\documentclass[pra,twocolumn,floatfix,aps]{revtex4}
\usepackage{amsmath}
\usepackage{makeidx}
\usepackage{amssymb}
\usepackage{graphicx}

\usepackage[usenames]{color}
\usepackage[normalem]{ulem}

\usepackage{hyperref}
\usepackage[T1]{fontenc} 

\begin{document}

\title{Quasi-one-dimensional soliton in a self-repulsive spin-orbit-coupled dipolar spin-half and spin-one condensates }

\author{S. K. Adhikari\footnote{sk.adhikari@unesp.br      \\  https://professores.ift.unesp.br/sk.adhikari/ }}

\affiliation{Instituto de F\'{\i}sica Te\'orica, Universidade Estadual Paulista - UNESP, 01.140-070 S\~ao Paulo, S\~ao Paulo, Brazil}
      

\date{\today}

\begin{abstract}

We study  the formation of  solitons 
in a  { uniform}   quasi-one-dimensional (quasi-1D)   spin-orbit (SO) coupled self-repulsive pseudo 
spin-half and spin-one dipolar   
Bose-Einstein condensates (BEC),   using  the mean-field Gross-Pitaevskii equation. The dipolar atoms are taken to be polarized  along the quasi-1D  $x$ direction. 
In  the pseudo spin-half case, for small SO-coupling,  one  can have dark-bright and bright-bright  solitons.
For large SO coupling,  the dark-bright and bright-bright  solitons may  acquire a spatially-periodic modulation in density; for certain values of contact interaction paramerers there is only     the normal  bright-bright soliton  without any spatially-periodic modulation in density.
In the spin-one anti-ferromagnetic case, for small SO coupling, one can have bright-bright-bright, dark-bright-dark,  and bright-dark-bright solitons;  and for large SO coupling, the 
 dark-bright-dark
 and  bright-dark-bright solitons are found to have a spatially-periodic modulation in density. 
 In the spin-one ferromagnetic case, for both small and large SO coupling, we  find only bright-bright-bright solitons.  All these solitons, specially those with a dark-soliton component, are dynamically stable as demonstrated  by real-time propagation using the converged stationary solution obtained by imaginary-time propagation as the initial state.

\end{abstract}

 \maketitle

\section{Introduction}

A bright soliton is a one-dimensional (1D) self-bound localized 
strongly-stable solitary wave that travel
with a constant velocity maintaining its shape due to
a cancellation of linear repulsion and nonlinear attraction. 
Such solitons have been found in a nonlinear medium \cite{nlm}, in water waves, in
nonlinear optics \cite{x2}, and in a Bose-Einstein condensate \cite{x1}  (BEC)
among others. Quasi-1D bright solitons have been created in a
BEC of $^7$Li \cite{r2a,r2b} and $^{85}$Rb \cite{r3} atoms, by a management of
the nonlinear attraction near a Feshbach resonance \cite{r4}, following the  suggestion of  a theoretical investigation  \cite{r5}, using  the mean-field Gross-Pitaevskii (GP) equation \cite{r6a,r6b}. 
There have been many studies of
  quasi-1D solitons in  a BEC \cite{38,39,40,41,42,43,44}.
 Dark solitons were also observed and studied in nonlinear optical fibers \cite{dsexpt1,dsexpt2,dsexpt3} and in 
a BEC of $^{87}$Rb atoms \cite{ds,dsy}.
A general dark soliton can also travel freely and is an envelope soliton having the form of a density dip with a phase
jump across its density minimum \cite{dsx}.
Bright (dark) solitons have a  maximum (minimum) of density 
at the center.  Although, the bright solitons in a BEC  have a long lifetime, the dark solitons    have a short lifetime and are usually unstable not only in a BEC \cite{ds,ds1,ds4} but also in a general nonlinear system \cite{ds3}. 

Not long  after the observation  \cite{BEC} of  condensates  of $^{85}$Rb, $^{87}$Rb,  $^7$Li  and  $^{23}$Na atoms in a laboratory,  Stenger et al.   observed and studied  a hyper-fine
spin-one three-component spinor BEC of $^{23}$Na atoms \cite{exptspinor}. 
There cannot be a usual spin-orbit (SO) coupling between two electrically neutral atoms in a BEC. The next step forward was the successful creation of an
artificial synthetic SO coupling  by  tuning   a  few 
Raman laser  beams which couple  two \cite{exptsob}  or three \cite{three} hyper-fine spin states of  electrically neutral 
spin-one $^{87}$Rb  \cite{exptso,thso,exptso2} and $^{23}$Na \cite{exptso3} atoms,   thus generating an SO-coupled pseudo spin-half or a spin-one BEC.
Multi-component SO-coupled spinor  BECs   possess distinct properties and can have different 
  excitations  \cite{kita1,kita2}, which are  not allowed  in a single-component scalar BEC.       For example, a quasi-1D \cite{quasi-1d}   soliton   stabilized  in 
a pseudo  spin-half or a spin-one  SO-coupled BEC  can have distinct spatial structure \cite{sg1,stripe}.

 The long-range non-local dipolar interaction  allows the formation of different types of   quasi-1D solitons in a BEC, which is not possible in the absence of a dipolar interaction.  After the experimental observation of dipolar BECs  of $^{52}$Cr \cite{1,2,6}, $^{166}$Er \cite{7}, $^{168}$Er \cite{8}, and  $^{164}$Dy
\cite{9,11}  atoms, with a large magnetic
dipole moment, we find it highly relevant to 
 perform  a detailed study of quasi-1D solitons in an SO-coupled  pseudo spin-half and    spin-one dipolar BEC.
   There have been a few previous studies on  different aspects of an   SO-coupled dipolar BEC \cite{sodip1,sodip2,sodip3,sodip4,sodip5,sodip6,sodip7,sodip8,sodip9,yuka,yuka1}  as well as on the formation of    a  quasi-1D \cite{sg1,sm1,quasi-1d,sm3},  quasi-two-dimensional (quasi-2D) \cite{quasi-2da,quasi-2db}  and quasi-three-dimensional
   \cite{hpu}  soliton
in an SO-coupled nondipolar BEC. These previous studies on   SO-coupled 
nondipolar BEC solitons were  performed on a {\it self-attractive} BEC with negative intraspecies  and interspecies  scattering lengths, where  solitons are possible even in the absence of an SO coupling. The formation of symbiotic 
bright solitons has also been studied in self-repulsive binary BEC mixtures \cite{r7a,r7b}
bound by interspecies attraction.

In this paper, using an appropriate mean-field model, we study the properties of different possible quasi-1D solitons in an SO-coupled dipolar pseudo  spin-half and spin-one self-repulsive BEC. Using the parameters of the  present model, no quasi-1D soliton is possible in the absence of an SO coupling and a dipolar interaction,  consistent with the term self-repulsive.  
 The dipolar atoms are taken to be polarized  along the quasi-1D $x$ direction in both pseudo spin-half and  spin-one cases.     This orientation of polarization leads to an attractive dipolar interaction necessary to bind a self-repulsive BEC to form a
soliton, which can freely move along the $x$ direction.

In the case of an SO-coupled dipolar pseudo spin-half BEC, we consider three distinct possibilities for  intraspecies and interspecies contact interactions: (i)  repulsive intraspecies and interspecies contact interactions, (ii) attractive intraspecies and repulsive interspecies interactions, and  (iii) repulsive intraspecies and attractive interspecies interactions. For the  above-mentioned  interaction (i),   for a small strength $\gamma$ of SO coupling, we have dark-bright, 
bright-bright and phase-separated  bright-bright solitons. [Unless specified to the contrary,  in the dark-bright (bright-bright)  soliton, the positions of 
the  maxima    of density of the bright and the minimum (maximum) of density of dark (bright) components coincide.]
In the case of  the above-mentioned  interaction (ii), for  a small $\gamma$,  we find   dark-bright and bright-bright solitons. 
For a large $\gamma$, for the above-mentioned interactions (i) and (ii) 
we have  bright-bright and dark-bright solitons, both with 
a spatially-periodic  stripe modulation in component density with a period  of $\pi/\gamma$, in addition to a bright-bright soliton without any modulation in density. 
In the case of the above-mentioned  interaction (iii),
both for  a small  and a large $\gamma$,  we find only bright-bright solitons.

For an SO-coupled dipolar spin-one BEC, the contact interaction among the three components is governed by  two parameters: $c_0$ and $c_2$ \cite{thso,exptsob}. Of these, 
a positive (negative) $c_2$ corresponds to an anti-ferromagnetic or polar  (ferromagnetic) BEC \cite{ku}. The properties of a spin-one ferromagnetic BEC could be   drastically different from a spin-one anti-ferromagnetic BEC \cite{ku}  and we consider these two distinct cases in this paper. In the case of an anti-ferromagnetic BEC ($c_2>0$), for a small strength $\gamma$ of SO coupling, one can have the dark-bright-dark and bright-dark-bright quasi-1D solitons, in addition to  partially phase-separated bright-bright-bright solitons. 
[Unless specified to the contrary,  in dark-bright-dark, bright-dark-bright and bright-bright-bright  solitons, the position of 
the  maximum    of density of the bright   component  coincides with the maximum (minimum) of density of other bright (dark) components.]
In the anti-ferromagnetic case,  for a large $\gamma$, one can have    dark-bright-dark and bright-dark-bright quasi-1D solitons with a spatially-periodic modulation in density with the period $\pi/ \gamma$. 
In the case of a ferromagnetic BEC ($c_2<0$),  for a small strength $\gamma$ of SO coupling, one can have the formation of overlapping bright-bright-bright solitons  and also  phase-separated  bright-bright-bright solitons. 
For a large $\gamma$, in this case, one can have   only   bright-bright-bright solitons without any  spatially-periodic modulation in density.

  The above-mentioned spatially-periodic modulation  
  \cite{14,exptso3,baym1,baym2,2020,stripe} in density   in a quasi-1D SO-coupled BEC soliton is
a consequence of the formation of a   supersolid \cite{s1,s2,s3,s4,s5}. A supersolid is a special quantum state of matter
with a rigid spatially-periodic crystalline structure \cite{54},
breaking continuous translational invariance, that flows
with zero viscosity as a superfluid breaking continuous
gauge invariance. Spatially-periodic supersolid stripe formation in a quasi-1D pseudo spin-half BEC  of $^{23}$Na atoms has been experimentally observed \cite{exptso3}.

 A dark soliton is an excited state and is unstable in general \cite{ds1}. However, the dark-bright soliton in  an SO-coupled pseudo spin-half  dipolar BEC and the dark-bright-dark and bright-dark-bright solitons in an SO-coupled spin-one dipolar BEC  are  dynamically  stable. We tested the dynamical stability of these solitons in a pseudo spin-half and a spin-one BEC by real-time propagation over a long time interval using the converged wave function obtained by imaginary-time propagation as the initial state.

 In Sec. \ref{i}  (Sec. \ref{ii}) we present the mean-field GP equation for the SO-coupled pseudo spin-half (spin-one)   dipolar BEC in dimensionless form appropriate for the present study of a quasi-1D soliton. In Sec. \ref{iii}  (Sec. \ref{iv}) we present an analytic study of the eigenfunctions  of the quasi-1D  solitons  in an SO-coupled pseudo spin-half   (spin-one) dipolar BEC.  We demonstrate the origin of bright-bright and dark-bright quasi-1D solitons in  the pseudo spin-half case.
 In the spin-one case, possible solitons could be   of  dark-bright-dark, bright-dark-bright or   bright-bright-bright  types.  In  Sec.  \ref{a}  (Sec. \ref{b}) we  present numerical results for 
 the pseudo spin-half  (spin-one) case.  
 In Sec. \ref{c}  we demonstrate the dynamical stability of these  solitons by real-time propagation over a long period of time after introducing a perturbation at time $t=0$.
  Finally, in Sec.  \ref{d} we present a brief summary of this investigation.  

\section{Mean-field model}

We consider a binary (pseudo spin-half)  or three-component (spin-one) SO-coupled  spinor BEC of $N$ atoms, each of mass { $m$},  under a
harmonic trap $V({\bf r})= m\omega_\rho^2(y^2+z^2)/2+ { m}\omega_x^2 x^2/2$ $(\omega_\rho \gg \omega_x)$ of angular frequency $\omega_x$ along the $x$ axis and 
and $\omega_\rho$     in the  quasi-2D $y$-$z$ plane.

{ The  atomic interaction between two atoms of  the ultradilute BEC has two parts:
the long-range anisotropic dipolar interaction $V_{\mathrm{dd}}(\bf R)$ and the $\delta$-function contact 
interaction $\delta({\bf R})$.}
 The  intraspecies ($V_j, j =1,2$) and 
interspecies ($V_{12}$)
interactions 
for two dipolar atoms, polarized along the $x$ axis, placed   at positions $\bf r$ and $\bf r'$ are given by \cite{1}
\begin{eqnarray}\label{intrapot} 
V_j({\bf R})&=&
\frac{\mu_0\mu^2}{4\pi}V_{\mathrm{dd}}({\mathbf R})+\frac{4\pi 
\hbar^2 a_j}{m}\delta({\bf R }),\\
 \label{interpot} 
V_{12}({\bf R})&=& \frac{\mu_0\mu^2}{4\pi}V_{\mathrm{dd}}({\mathbf R})+
\frac{4\pi \hbar^2 a_{12}}{m}\delta({\bf R}),
\end{eqnarray}
\begin{eqnarray}
 \label{dp}
V_{\mathrm{dd}}({\mathbf R})  &=& 
\frac{1-3\cos^2\theta}{{\mathbf R}^3},
     \end{eqnarray}
where $\bf R \equiv (r-r')$ is the position vector joining the two atoms  at $\bf r$ and $\bf r '$,      $\mu_0$ is the permeability of free space, $\mu$  is the magnetic dipole moment of each atom,
$\theta$ is the angle made by the vector ${\bf R}$ with the polarization 
$x$ direction,  $a_j$ is the intraspecies scattering length { of component $j$}, and $a_{12}$  is the interspecies  scattering length. 
To compare the dipolar and contact interactions, the  
dipolar interaction  is  expressed in terms of the dipolar length, defined by \cite{1}
\begin{align}
  a_{\mathrm{dd}}\equiv 
\frac{\mu_0\mu^2m}{12\pi \hbar ^2}. 
\end{align}
The dipolar length measures the strength of dipolar interaction in the same way as the scattering length measures the strength of contact interaction in an ultra-dilute BEC.
As the different  components have the same type of atoms with the same magnetic dipole moment, both intraspecies and the interspecies dipolar interactions and the two (pseudo spin-half BEC) or three (spin-one BEC) possible   dipolar lengths are equal.  

\subsection{SO-coupled spin-half dipolar GP equation}

\label{i}

 {
 In cold atoms an artificial synthetic  SO coupling is created through the Zeeman interaction  $-{\boldsymbol \mu}\cdot \bf B$
 between the magnetic  moment $\boldsymbol \mu$ of an atom, proportional to its spin $\boldsymbol \sigma$, and a magnetic field ${\bf B}\sim 
E_0 (-p_y\hat x+ p_x \hat y)$ generated  in
the   moving frame \cite{exptso,ku} of the atom moving with momentum $\bf p$, due to  an external  static  electric
field $E=E_0\hat z$ in the lab frame.   This leads to an SO coupling $-{\boldsymbol \mu}\cdot {\bf B }\sim (\sigma_xp_y -\sigma_yp_x).$ 
The Pauli spin matrices  $\sigma_x,$  $\sigma_y$ and $\sigma_z$  are 
\begin{eqnarray}\label{choice}
\sigma_x=\begin{pmatrix}
0 & 1  \\
1 & 0 
\end{pmatrix}, \quad
\sigma_y=
\begin{pmatrix}
0 & -\mbox{i}  \\
\mbox{i} & 0 
\end{pmatrix},\quad
\sigma_z=
\begin{pmatrix}
1 & 0 \\
0 & -1
\end{pmatrix},
\end{eqnarray}
 where ${\mbox{i}=\sqrt{-1}}.$ For a strict 1D system moving  along  $x$ direction,  $p_y=0$  and we can generate an SO coupling of the type $\sigma_y p_x$.
Similarly, it is also possible to create  an SO coupling of  the type $\sigma_x p_x$.
} 
 The single particle Hamiltonian of the SO-coupled  condensate can be written as
 \cite{exptso} 
\begin{align} 
H_0 =-\frac { \hbar^2}{2 m}  \nabla_{\bold r}^2 +  V({\bold r})+\gamma p_x \eta,
\end{align}
where ${\bold r}\equiv \{x,y,z   \}$,  ${\boldsymbol \rho} \equiv \{ y,z\}$,
 {$\nabla_{\bold r}^2=
(\partial^2/\partial x^2+\partial^2/\partial y^2+\partial ^2 /\partial z^2) \equiv (\partial_x^2+\partial_y^2+\partial_z^2)$}, $p_x=-{\mbox i}\hbar \partial_x$ is the momentum along the $x$ axes, 
$\gamma$ is the strength of the SO-coupling interaction.  We  will consider the two possibilities of SO coupling given by $\eta = \sigma_x$ and $\eta =\sigma_y.$ Of these two types of SO couplings, the choice $\gamma \sigma_y  p_x$  corresponds to an equal mixture of Dresselhaus \cite{dre} and Rashba \cite{ras} SO couplings  and was realized in the  pioneering  experiments  on SO coupling in a pseudo spin-half BEC of $^{87}$Rb \cite{exptso} and $^{23}$Na \cite{exptso3} atoms. We will not consider the choice $\eta=\sigma_z$, as that choice does not permit any off-diagonal coupling 
and hence does not 
 lead to the variety of solitons found with the  choices  (\ref{choice}) illustrated in Figs. \ref{fig1}-\ref{fig3} and leads to only bright-bright solitons (result not elaborated in  this paper).

This dipolar pseudo spin-half binary BEC, without any synthetic SO coupling,  of two components $j=1,2$  is described by the following set of equations \cite{1,bao}
\begin{eqnarray}
\label{Eq1}
{\mbox i} \hbar \partial_t  \phi_j({\bf r},t)   &=&\frac{\hbar^2}{m}
\Big [ -\frac{1}{2}(\partial_x^2+\partial_y^2+\partial_z^2) \nonumber \\&+& \frac{m^2}{2\hbar^2}  
(\omega_\rho ^2\rho^2+\omega_x^2 x^2 )
\nonumber\\ 
&+& 4\pi {a}_j N_j \vert \phi_j({\bf r},t) \vert^2
+4\pi  {a}_{12} N_k \vert \phi_k({\bf r},t) \vert^2 \nonumber
\\ 
&+&{3}a_{\mathrm{dd}} N_j \int  V_{\mathrm{dd}} ({\mathbf R})\vert\phi_j({\mathbf r'},t)\vert^2 d{\mathbf r}' \nonumber \\
&+&3a_{\mathrm{dd}} N_k \int  V_{\mathrm{dd}} ({\mathbf R})\vert\phi_k
({\mathbf r'},t)\vert^2 d{\mathbf r}' 
\Big] 
 \phi_j({\bf r},t),
 \nonumber \\
 &j &\ne k=1,2,
\end{eqnarray}
where  $\partial_t \equiv \partial /\partial t $, $N_j$ and $N_k$ are the number of atoms in the two components $j$ and $k$, respectively ($N=N_1+N_2$).
Here $j=1$ is the  spin-up component and $j=2$ is the 
spin-down component.
Equations (\ref{Eq1}) can be cast in the following dimensionless form if we scale lengths in units of the  harmonic oscillator length $l_0=\sqrt{\hbar/m\omega_\rho}$, time units of $t_0=1/\omega_\rho$,  %
density $|\phi_j|^2$  in units of  $l_0^{-3}$, 
and  energy in units of $\hbar \omega_\rho$ 
\begin{align}
{\mbox i} \partial_t   \phi_j({\bf r},t)&=
{\Big [}  -\frac{1}{2}(\partial_x^2+\partial_y^2+\partial_z^2)+ \frac{1}{2}  
\left(\rho^2+\frac{\omega_x ^2}{\omega_\rho^2} x^2 \right)
\nonumber\\ &
+ {4\pi }{a}_j N_j \vert \phi_j({\bf r},t) \vert^2
+{4\pi } {a}_{12} N_k \vert \phi_k({\bf r},t) \vert^2
\nonumber 
\\  
&+{3}a_{\mathrm{dd}} N_j \int  V_{\mathrm{dd}} ({\mathbf R})\vert\phi_j({\mathbf r'},t)\vert^2 d{\mathbf r}' \nonumber \\
&+{3}a_{\mathrm{dd}} N_k \int  V_{\mathrm{dd}} ({\mathbf R})\vert\phi_k({\mathbf r'},t)\vert^2 d{\mathbf r}'
\Big]  \phi_j({\bf r},t), \nonumber \\
&  j\ne k=1,2.
\label{eq31}
\end{align}
Without any risk of confusion,
here and in the following we are using the same symbols to denote the scaled dimensionless and unscaled variables.

We consider a quasi-1D  pseudo spin-half dipolar BEC with a strong trap in the $y$-$z$ plane.  We assume that the dynamics of the BEC in the $y$-$z$ plane  is frozen in the ground state 
\begin{equation}
\psi_B(\rho) ={\pi}^{-1/2} e^{-\rho^2 /2} ,  \quad \rho^2=y^2 +z^2,
\end{equation}
{and the relevant dynamics of the BEC will be confined along the polarization $x$ direction. In this case it is possible to integrate out the $y$ and $z$ variables and write a set of coupled equations for the relevant dynamics along the $x$ direction.   }
The component wave function of the BEC can be written as 
\begin{equation}\label{model}
\phi_j({\bf r},t)= \psi_B(\rho)\times \psi_j (x,t),
\end{equation}
where the wave function $\psi_j(x,t)$ describes the relevant dynamics of the  quasi-1D BEC confined along the $x$ direction. {To obtain
 the relevant dynamics,  we substitute  Eq. (\ref{model})  into 
Eq. (\ref{eq31}), multiply the resultant equation by $\psi_B(\rho)$  and integrate  over 
 ${\boldsymbol \rho }\equiv \{y,z \}$  \cite{2d-3d}.    To avoid the difficulty associated with the integration over the divergent dipolar interaction in configuration space, this integral is conveniently evaluated in the momentum space.
 This procedure leads to the}  
 following set of quasi-1D  GP equations   for the two  components  \cite{bec2015,bao} of the pseudo spin-half SO-coupled dipolar BEC soliton
\begin{align}
\label{EQ1} 
{\mbox i} \partial_t \psi_{j}(x,t)&= \Big[ - \frac{1}{2}  \partial_x^2
+c_0 n_j({x},t)  + c_2n_k(x,t) \nonumber \\
 &
+ d\sum_{j=1,2} s_j(x,t)  \Big] \psi_{j}(x,t)  + [{ \gamma} p_x  \eta \psi(x,t)]_j,
 \\  
 s_j(x,t)&=  \int \frac{d{ k}_x}{2\pi} e^{-{\mbox i} {k}_x x}
\widetilde n_j({k}_x,t)h_{1D}\left(\frac{k_x }{\sqrt {2 }}  \right) \,  , \label{sj} \\
h_{1D}\left(\frac{k_x }{\sqrt {2 }}  \right) &=   \frac{1}{(2\pi)^2}\int d {\bf k}_\rho\left(
\frac{3k_x^2}{{\bf k}^2}-1\right)   |\widetilde n({{\bf k}_\rho})| ^2 \nonumber \\
&=  \frac{1}{2\pi} \int_0 ^\infty du \left[  \frac{3k_x^2}{2u+3k_x^2}-1 \right] e^{-u} ,
\end{align}
{where $\widetilde n_j({k}_x,t)$ and $\widetilde  n({\bf k}_\rho)$ are the Fourier transformations of densities $|\psi_j(x,t) |^2$ and $\widetilde  n({\bf k}_\rho) $, respectively, and are given by}
\begin{align}
\widetilde n_j({k}_x,t)&=\int_{-\infty}^{\infty}  e^{{\mbox i}k_x x} |\psi_j(x,t) |^2 dx ,\\
\widetilde  n({\bf k}_\rho) &= \int   e^{i{\bf k} _\rho \cdot {\boldsymbol  \rho}}| \psi_B(\rho) |^2 d
{\boldsymbol \rho}=  e^{-k_\rho ^2/4},
\end{align}
where $k_\rho= \sqrt{k_y^2 +k_z  ^ 2},$
and we have included the SO-coupling term  in Eq. (\ref{EQ1}) and  assumed that $a_1=a_2\equiv a$.  
The   dipolar intraspecies or interspecies nonlinearity $d=4\pi a_{\mathrm{dd}}N,$
intraspecies contact nonlinearity $c_0 = 2N a,$  interspecies contact nonlinearity $c_2=2N a_{12}$.  
The chosen interaction parameters will be such that there could be no solitons in the absence of SO coupling and dipolar interaction.
Here     
 $n_j(x,t) = |\psi_j({x,t})|^2, j=1,2$ are the densities of the two components, and $n(x,t)= \sum_j n_j({x},t)$  the total density. In Eq. (\ref{EQ1}), for $\eta =\sigma_x$,
 \begin{equation}
  [{ \gamma} p_x  \eta \psi(x,t)]_j=-{\mbox{i}} \gamma \partial_x \psi_k(x,t),  \quad  j\ne k,
  \end{equation} and for $\eta =\sigma_y,$
  \begin{equation}
 [{ \gamma} p_x  \eta \psi(x,t)]_j=(-1)^k \gamma \partial_x \psi_2(x,t),  \quad  j\ne k,
 \end{equation}
 $j, k=1,2$.
 In this study of quasi-1D solitons along the $x$ direction we have dropped the harmonic oscillator trapping  potential from Eq. (\ref{EQ1}).
The normalization condition on  the total density is 
\begin{align}\label{noma}
 {\textstyle \int} _{-\infty}^{\infty} n (x,t)\, dx = 1.
\end{align}    The SO coupling allows a transfer of atoms from one component to another so that there is no condition of  normalization on the individual components. The disbalance between the spin-up and spin-down components is measured by magnetization ${\cal M}$, defined by 
\begin{align}
{\cal M}=\frac{\int_{-\infty}^{\infty}dx{[n_1(x)-n_2(x)]}}{{\int_{-\infty}^{\infty}}dx{[n_1(x)+n_2(x)]}}.
\end{align}
{Imaginary-time propagation does not conserve normalization automatically. 
In each time iteration the total normalization is adjusted to unity, and the normalization of the individual components is allowed to vary freely with time  propagation.  
 Imaginary-time propagation settles the final  converged density of the components and the final  magnetization ${\cal M}$. }
 
\subsection{SO-coupled spin-one dipolar GP equation}

\label{ii}
 
 In  this case the mean-field   equations are  similar to those of a pseudo spin-half SO-coupled dipolar BEC  and  we  highlight the differences  from the 
 pseudo spin-half case.  Here the SO-coupled  spinor BEC has  three components $j=1,2,3$ in contrast to a pseudo spin-half BEC with two components $j=1,2$.  {  Again, $j=1$ represents spin-up  component and $j=2$  represents the spin-down component. We  will consider the  same SO-coupling interaction in this case as in the case of an  SO-coupled pseudo spin-half dipolar BEC in Sec. \ref{i}. }
  The  single-particle Hamiltonian is now given by 
 \begin{align} 
H_0 &=-\frac { \hbar^2}{2 m} (\partial_x^2+\partial_y^2+\partial_z^2) +  V({\bold r})+\gamma p_x \eta.
\end{align}
However, { the spin operator now has three spin components and }
the spin-one matrices are  $3\times 3$  matrices. For $\eta$ we now consider {the  two possibilities $\eta=\Sigma_x$ and $\Sigma_y$, where}
\begin{align}\label{choice1}
\Sigma_x & =\frac{1}{\sqrt 2}\begin{pmatrix}
0 & 1  & 0\\
1 & 0 & 1\\
0 & 1 & 0 \\
\end{pmatrix}, \quad 
\Sigma_y  =\frac{1}{\sqrt 2}\begin{pmatrix}
0 & -\mbox{i}  & 0\\
\mbox{i} & 0 & -\mbox{i} \\
0 &  \mbox{i}& 0 \\
\end{pmatrix},
\end{align}
corresponding to the SO coupling $\gamma p_x  \Sigma_x$  and $\gamma p_x  \Sigma_y$, respectively.  We will not consider the possibility 
\begin{align}\label{avt}
\Sigma_z & = \begin{pmatrix}
1 & 0  & 0\\
0 & 0 & 0\\
0 & 0 & -1 \\
\end{pmatrix}, 
\end{align}
where there is no off-diagonal coupling. 
Consequently, the  choice  (\ref{avt}) does not lead to the variety of solitons found with the  choices  (\ref{choice1}), illustrated in Figs. \ref{fig4}-\ref{fig5}, and leads only to bright-bright-bright solitons. 
 
 The quasi-1D equations for the three components of   a SO-coupled spin-one 
 BEC are well known \cite{sg1,ku}, {e.g., Eqs. (\ref{eq1})-(\ref{eq3}) with $d=\gamma=0$}. For a dipolar BEC {($d\neq 0$)} the dipole-dipole interaction terms are introduced  {in these equations} as in the case of a pseudo spin-half system, viz.  Eq. (\ref{EQ1}).  The resultant quasi-1D equations  for a dipolar spin-one  SO-coupled BEC of $N$ atoms can be obtained by combining the same of a dipolar BEC \cite{bec2015} with that of a spin-one BEC \cite{xy1,xy2,sg1} and 
 are as follows
 \begin{align}
\label{eq1} 
{\mbox i} \partial_t \psi_1(x,t)&= \Big[ - \frac{  \partial_x^2}{2} 
+c_0 n({x},t) +c_2\{n_1(x,t)+n_2(x,t) \nonumber \\
 &-n_3(x,t)\}
+ d\sum_{j=1}^3 s_j(x,t)  \Big] \psi_{1}(x,t) 
\nonumber \\
&+ c_2\psi_2^2(x,t) \psi_3^*(x,t)+[\gamma p_x  \eta \psi(x,t)]_1,  
\\ \label{eq2} 
{\mbox i} \partial_t \psi_2(x,t)&= \Big[ - \frac{  \partial_x^2}{2} 
+c_0 n({x},t) +c_2\{n_1(x,t)+n_3(x,t)\} \nonumber \\
 &+
 d\sum_{j=1}^3 s_j(x,t)  \Big] \psi_2(x,t) +2c_2\psi_1(x,t) \psi_3(x,t) 
\nonumber \\&\times  \psi_2^ *(x,t)+ [{ \gamma} p_x  \eta \psi(x,t)]_2,  
\\ \label{eq3} 
{\mbox i} \partial_t \psi_3(x,t)&= \Big[ - \frac{  \partial_x^2}{2} 
+c_0 n({x},t) +c_2\{n_3(x,t)+n_2(x,t) \nonumber 
\\ 
 &-n_1(x,t)\}
+ d\sum_{j=1}^3 s_j(x,t)  \Big] \psi_{3}(x,t) 
\nonumber \\&+ c_2\psi_2^2(x,t) \psi_1^*(x,t)+ [{ \gamma} p_x  \eta \psi(x,t)]_3,  
\end{align}
{where we have introduced the SO-coupling terms $ [{ \gamma} p_x  \eta \psi(x,t)]_j, j=1,2,3.$ }
Here the spin-one interaction parameters are \cite{sg1,ku} $c_0=2N(a_0+2a_2)/l_0$,  $c_2=2N(a_2-a_0)/l_0$, $a_0$ and $a_2$ are the $s$-wave
scattering lengths of two spin-one atoms in the total spin   0 and 2 channels. For a ferromagnetic BEC  $c_2<0$ and for an anti-ferromagnetic BEC  $c_2>0$ \cite{ku}. 
In this study we will consider $c_0 >0$  and  a small $|c_2|$, including both ferromagnetic and anti-ferromagnetic domains.
 {For this choice of $c_0$ and $c_2$,} there cannot be any solitons in the absence of the dipolar interaction and SO-coupling;    we  will term such a BEC as self-repulsive indicating that the net contact interaction is repulsive, { although $c_2$ can have a small negative value in the ferromagnetic domain ($|c_0|>|c_2|).$}  For a spin-one BEC, the background scattering lengths of $^{87}$Rb and $^{23}$Na atoms fall in the ferromagnetic \cite{24, 25a,25b} and anti-ferromagnetic \cite{26} domains,
respectively. In Eqs. (\ref{eq1})-(\ref{eq3}),
for $\eta =\Sigma_x$, 
\begin{align}
[{ \gamma} p_x  \eta \psi(x,t)]_j &=-\frac{\mbox i}{\sqrt 2}\gamma  \partial_x \psi_2(x,t), \quad j=1,3,  \\  [{ \gamma} p_x  \eta \psi(x,t)]_2 &=-\frac{\mbox i}{\sqrt 2}\gamma   
[\partial_x \psi_1(x,t)+\partial_x \psi_3(x,t)], 
\end{align} and for   $\eta =\Sigma_y$,
\begin{align}
 [{ \gamma} p_x  \eta \psi(x,t)]_j&= \frac{(-1)^j}{\sqrt 2}\gamma  \partial_x \psi_2(x,t), \quad  j=1,3,   \\ [{ \gamma} p_x  \eta \psi(x,t)]_2& =\frac{1}{\sqrt 2}\gamma  
[\partial_x \psi_1(x,t)-\partial_x \psi_3(x,t)].
\end{align}
 {In  this case also the normalization of individual components is not conserved  and only the total normalization of the three components is conserved.} The 
 magnetization
\begin{align}
{\cal M}=\frac{\int_{-\infty}^{\infty}dx{[n_1(x)-n_3(x)]}}{{\int_{-\infty}^{\infty}}dx{[n_1(x)+n_2(x)+n_3(x)]}},
\end{align}
{ gives a measure of the disbalance between the spin-up and spin-down components.
 }
 
 \section{Analytical results}
 
 \subsection{SO-coupled spin-half GP equation}
 
 \label{iii}

 Equations  (\ref{EQ1})  can be derived from the energy functional
 \begin{eqnarray}\label{ehalf}
 E[\psi_j] & =&\textstyle \frac{1}{2} \int_{-\infty}^{\infty}  dx
  {\Huge[}  \sum_{j =1}^2 |\partial _x \psi_j|^2+c_0(n_1^2+n_2^2)  \nonumber \\ &+&2c_2n_1n_2+ d( s_1+s_2)n-2{\mathrm i}
 \psi^T  \gamma \partial_x \eta \psi  \Huge]\, ,
 \end{eqnarray}
 where $\psi^T = (\psi_1,\psi_2),$  {using the variational rule \cite{BEC}
 \begin{align}\label{rule}
{\mbox  i }\partial_t \psi_j  = \frac{\delta E [\psi_j]}{\delta \psi_j^*}.
 \end{align}
}
 
Many  properties of a pseudo spin-half, nondipolar or dipolar,    SO-coupled uniform BEC  can be understood from a consideration  of the eigenfunction-eigenvalue problem of the linear Hamiltonian, setting all nondipolar and dipolar nonlinear interactions and the confining trap  to zero [$c_0=c_2=d=V({x})=0$] \cite{sg1}. The quasi-1D  stationary wave function of the linear single-particle trap-less   Hamiltonian  corresponding to Eq. (\ref{EQ1}) is 
\begin{equation}\label{abc1}
H_{0}^{1D}= -\textstyle \frac{1}{2}\partial_x ^2-{\mbox i}\gamma  \partial_x \eta,
\end{equation}
which for $\eta =\sigma_x$  satisfies the following  eigenfunction-eigenvalue problem
for a stationary state
\begin{eqnarray}
\begin{pmatrix}\label{spf1}
-\textstyle \frac{1}{2}\partial^2_x & -{\mbox i}\gamma \partial_x  \\
-{\mbox i}\gamma\partial_x  & -\textstyle \frac{1}{2}\partial_x^2
\end{pmatrix}   \begin{pmatrix} \psi_1 (x)\\ \psi_2(x) \end{pmatrix}
=  {\cal E}   \begin{pmatrix} \psi_1 (x)\\ \psi_2 (x)\end{pmatrix},
\end{eqnarray}
which has  the following two degenerate  eigenfunctions with the eigenvalue ${\cal E}=-\gamma^2/2$:
\begin{align}\label{ev32}
  \begin{pmatrix} \psi_1 (x)\\ \psi_2(x) \end{pmatrix}
=  \begin{pmatrix} \cos (\gamma x)\\  -{\mathrm i}\sin(\gamma x) \end{pmatrix}, \quad  \begin{pmatrix} \psi_1 (x)\\ \psi_2(x) \end{pmatrix}
=   \begin{pmatrix}  \sin (\gamma x)\\ {\mathrm i}\cos(\gamma x)\end{pmatrix}.
\end{align}
 For   $\eta =\sigma_y$  we  have the following eigenfunction-eigenvalue problem
\begin{eqnarray}
\begin{pmatrix}\label{spf2}
-\textstyle \frac{1}{2}\partial^2_x & -\gamma \partial_x  \\
+\gamma\partial_x  & -\textstyle \frac{1}{2}\partial^2_x
\end{pmatrix}   \begin{pmatrix} \psi_1(x) \\ \psi_2 (x)\end{pmatrix}
=  {\cal E}   \begin{pmatrix} \psi_1(x)\\ \psi_2(x) \end{pmatrix}, 
\end{eqnarray}
which has  the following two degenerate  eigenfunctions with the eigenvalue ${\cal E}=-\gamma^2/2$:
\begin{align}\label{ev31}
  \begin{pmatrix} \psi_1 (x)\\ \psi_2(x) \end{pmatrix}
=  \begin{pmatrix} \cos (\gamma x)\\  \sin(\gamma x) \end{pmatrix} , \quad  \begin{pmatrix} \psi_1 (x)\\ \psi_2(x) \end{pmatrix}
=   \begin{pmatrix}  \sin (\gamma x)\\ -\cos(\gamma x)\end{pmatrix}.
\end{align}
  In both cases, { $\eta=\sigma_x, \sigma_y,$ we find from Eqs. (\ref{ev32}) and (\ref{ev31}) that} the  energy and the densities are the same, although the wave functions are different.   The density of the two components has  a  spatially-periodic modulation, given by  $\sin^2 (\gamma x)$ and $\cos^2(\gamma x)$, along the quasi-1D $x$ direction with a period of $\pi/\gamma$; 
  however, the total density  does not have  this modulation {and $|\psi_1(x)|^2+ |\psi_2(x)|^2=1$}. The two component densities  have a maximum  and a minimum at the origin $x=0$ leading to a  dark-bright soliton in the uniform system.

 \subsection{SO-coupled spin-one GP equation}
 
 \label{iv}

 The time-independent version of Eqs. (\ref{eq1})-(\ref{eq3}), appropriate for the stationary solutions,  can be derived from the energy functional
 \begin{align}
 E[\psi_j]&=\textstyle \frac{1}{2}\int _{-\infty}^ {\infty}dx\big[\sum_{j=1}^3|\partial_x \psi_j|^2
 +c_0 n^2+c_2\{n_1^2+n_3^2\nonumber \\&+2(n_1n_2+n_3n_2-n_1n_3+\psi_3^*\psi_2^2 \psi_1^ *
 +\psi_3\psi_2^{* 2}  \psi_1)\}\nonumber \\
 & + d(s_1 +s_2+s_3)n-2{\mathrm i} \psi^T \gamma \partial_x \eta \psi
 \big], 
 \end{align}{
 using the variational rule  (\ref{rule})}, where now $\psi^T = (\psi_1,\psi_2,\psi_3)$.

Again, 
many  properties of a   spin-one    SO-coupled uniform BEC  can be understood from a consideration  of the eigenfunction-eigenvalue problem of the linear Hamiltonian, setting all nondipolar and dipolar nonlinear interactions and the confining trap  to zero [$c_0=c_2=d=V(x)=0$] \cite{sg1}. The quasi-1D  stationary wave function of the linear single-particle trap-less   Hamiltonian  corresponding to Eqs. (\ref{eq1})-(\ref{eq3})
 \begin{equation}\label{abc}
H_{0}^{1D}= -\textstyle \frac{1}{2}\partial_x ^2-{\mbox i}\gamma  \partial_x \eta,
\end{equation}
which, for $\eta =\Sigma_x,$  satisfies the following  eigenfunction-eigenvalue problem
\begin{align}
\begin{pmatrix}\label{spf3}
-\textstyle \frac{1}{2}\partial^2_x & -\frac{\mbox i}{\sqrt 2}\gamma \partial_x & 0\\
\textstyle
-\frac{\mbox i}{\sqrt 2}\gamma\partial_x  & -\textstyle \frac{1}{2}\partial_x^2& -\frac{\mbox i}{\sqrt 2}\gamma \partial_x\\0 & -\frac{\mbox i}{\sqrt 2}\gamma \partial_x &  -\textstyle \frac{1}{2}\partial^2_x
\end{pmatrix}   \begin{pmatrix} \psi_1 (x)\\ \psi_2(x) \\ \psi_3(x) \end{pmatrix}
=  {\cal E}   \begin{pmatrix} \psi_1 (x)\\ \psi_2 (x) \\  \psi_3(x) \end{pmatrix},\nonumber \\
\end{align}
which has the following two linearly independent  degenerate eigenfunctions with energy ${\cal E}= -\gamma^2/2$:
\begin{align}\label{ev1}
  \begin{pmatrix} \psi_1 (x)\\ \psi_2(x) \\ \psi_3(x) \end{pmatrix}
=   \frac{1}{\sqrt 2} \begin{pmatrix} \cos (\gamma x)\\ -{\mathrm i}\sqrt 2 \sin(\gamma x)\\  \cos(\gamma x) \end{pmatrix}, \\   \begin{pmatrix} \psi_1 (x)\\ \psi_2(x) \\ \psi_3(x) \end{pmatrix}
=   \frac{1}{\sqrt 2} \begin{pmatrix}  \sin (\gamma x)\\ {\mathrm i} \sqrt 2 \cos(\gamma x)\\    \sin (\gamma x \end{pmatrix}.
\label{ev2}
\end{align}

  For   $\eta =\Sigma_y,$  we  have the following eigenfunction-eigenvalue problem
\begin{align}
\begin{pmatrix}\label{spf4}
-\textstyle \frac{1}{2}\partial^2_x & -\frac{1}{\sqrt 2}\gamma \partial_x & 0\\
\textstyle
\frac{1}{\sqrt 2}\gamma\partial_x  & -\textstyle \frac{1}{2}\partial_x^2& -\frac{1}{\sqrt 2}\gamma \partial_x\\
0 & \frac{1}{\sqrt 2}\gamma \partial_x &  \textstyle -\frac{1}{2}\partial^2_x
\end{pmatrix}   \begin{pmatrix} \psi_1 (x)\\ \psi_2(x) \\ \psi_3(x) \end{pmatrix}
=  {\cal E}   \begin{pmatrix} \psi_1 (x)\\ \psi_2 (x) \\  \psi_3(x) \end{pmatrix},\nonumber 
\\
\end{align}
which has  the following two degenerate  eigenfunctions with eigenvalue ${\cal E}=-\gamma^2/2$:
\begin{align}\label{ev3}
  \begin{pmatrix} \psi_1 (x)\\ \psi_2(x) \\ \psi_3(x) \end{pmatrix}
=   \frac{1}{\sqrt 2} \begin{pmatrix} \cos (\gamma x)\\ \sqrt 2 \sin(\gamma x)\\          -\cos(\gamma x) \end{pmatrix}, \\   \begin{pmatrix} \psi_1 (x)\\ \psi_2(x) \\ \psi_3(x) \end{pmatrix}
=   \frac{1}{\sqrt 2} \begin{pmatrix}  \sin (\gamma x)\\ - \sqrt 2 \cos(\gamma x)\\    -\sin (\gamma x) \end{pmatrix}.\label{ev4}
\end{align}

The densities  corresponding to eigenfunctions  (\ref{ev1}) and 
(\ref{ev2}) are  identical to the same of eigenfunctions  (\ref{ev3}) and 
(\ref{ev4}).   These densities correspond to a spatially periodic sinusoidal stripe  along the $x$ axis with a period of $\pi/\gamma$;  the  stripe pattern of components $j=1$ and 3 is always the same and that of component $j=2$ is always displaced with respect to the former.
  The  former may have a maximum  (minimum)  at the origin $x=0$,  where  the latter has a minimum (maximum)
leading to a bright-dark-bright  or a dark-bright-dark soliton, respectively,  in a uniform system. The total density of the three components do not  have any modulation in density and is uniform in nature: { $|\psi_1(x)|^2+|\psi_2(x)|^2+|\psi_3(x)|^2=1$.  The solutions presented in Secs. \ref{iii} and \ref{iv} of the linear equations are not localized.   We
will see in Sec. \ref{v}, the localized  solitons of  the full nonlinear equations have the same modulation in density.}

\section{Numerical Result}
 \label{v}

To solve the binary equations   (\ref{EQ1})  for a quasi-1D SO-coupled pseudo spin-half dipolar BEC  and three-component equations   (\ref{eq1})-(\ref{eq3})   for a quasi-1D SO-coupled spin-one dipolar BEC numerically, we propagate
these  in time by the split-time-step Crank-Nicolson discretization scheme \cite{bec2009}.
{  In this study of solitons in an SO-coupled dipolar pseudo spin-half or spin-one BEC, all the (contact and dipolar)  interaction parameters  will be taken to be very small in magnitude, so that the system  remains in the very weak-coupling limit \cite{BEC}, where     beyond-mean-field effects are negligibly small and not of concern, even in the presence of dipolar interaction.  On the other extreme of very strong-coupling limit of droplet formation in  a  dipolar BEC,  quantum-fluctuation effects are highly relevant and  are incorporated in the mean-field model through the  higher-order Lee-Huang-Yang interaction \cite{xyz}, or one can also employ  different beyond-mean-field  formulations.}
As we require to solve the GP equation in the presence of  both SO-coupling and dipolar  interaction,  we  needed  to combine the Open Multiprocessing Programs  for solving the 
dipolar \cite{bec2023} and SO-coupled \cite{bec2021} GP equations.
 We employ the  space
step  $dx=0.1$  and the  time step  $dt=dx^2\times  0.1$ for  imaginary-time propagation
and the  time step  $dt=dx^2\times  0.025$ for real-time propagation.   { In all calculations of  stationary states, we  use  imaginary-time propagation with the 
conservation of total normalization  
during time propagation, which 
finds the lowest-energy solution of each type.  The real-time propagation is used to demonstrate the dynamical stability of the solitons.

In the presence of SO coupling, the transfer of atoms from one spin component to another is allowed \cite{ku} and 
 the number of atoms in each component   is not separately conserved. Consequently, 
 the magnetization  ${\cal M}$ is
not a good quantum number in the presence of SO coupling. Hence the magnetization as well as the  normalization of individual components  is not conserved during imaginary-time propagation.   As  the total 
number of atoms in all the   components is conserved,  the conservation of total normalization is enforced in each time step of imaginary-time propagation. In  addition, if necessary, the 
conservation of magnetization in each time step can be   enforced as in Ref. \cite{sg1}. 
 Some of the present solitons are possible for a zero as well as  a nonzero value of magnetization {$\cal M$}; in those cases we  present the  soliton with zero magnetization in  this paper. For a dark-bright soliton in the case of  a  pseudo spin-half dipolar BEC, the magnetization is intrinsically non-zero; in that case we illustrate the ground state with ${\cal} M\ne 0$ obtained by  the imaginary-time  propagation.

\subsection{ Quasi-1D SO-coupled    spin-half self-repulsive  dipolar BEC soliton}
 
\label{a}  

We study the formation of different types of solitons in a quasi-1D SO-coupled   pseudo spin-half self-repulsive   dipolar  BEC.  
In  this system we will choose the intraspecies and interspecies scattering lengths in such a way  that  no solitons are possible in the absence of a  dipolar interaction and an SO coupling. Both a dipolar interaction  \cite{bec2015}  and an SO coupling \cite{quasi-2da} contribute to an attraction in the system and these help to form a bound soliton.  We will consider three possibilities  of intraspecies and interspecies interactions:  repulsive intraspecies and interspecies contact interactions,  attractive intraspecies and repulsive interspecies contact  interactions, and repulsive  intraspecies  and attractive interspecies contact  interactions.

\begin{figure}[!t] 
\centering 
\includegraphics[width= .49\linewidth]{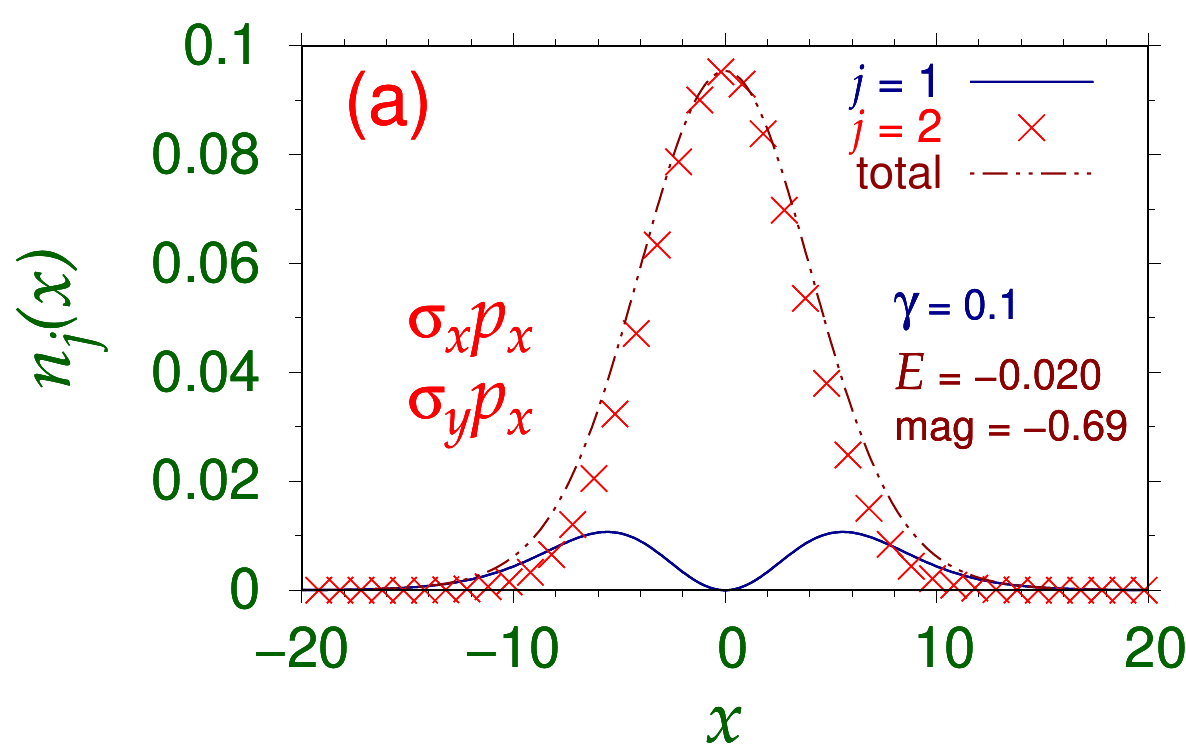}  
\includegraphics[width= .49\linewidth]{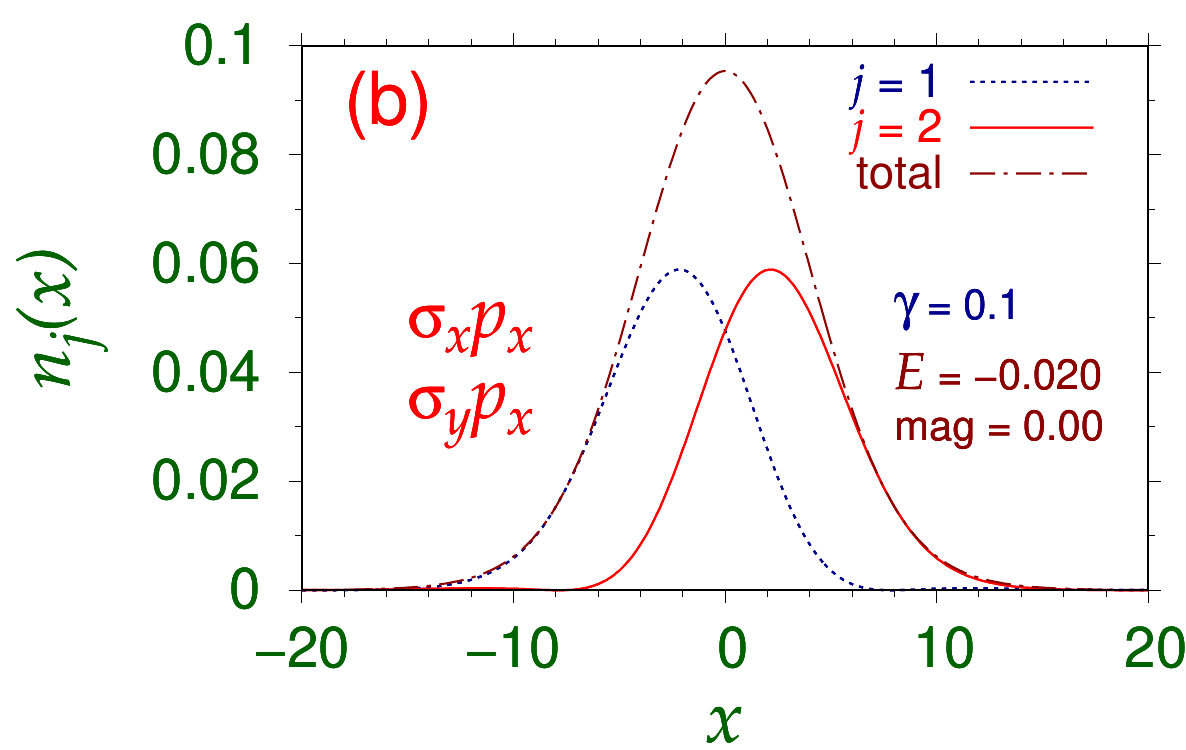}  
\includegraphics[width= .49\linewidth]{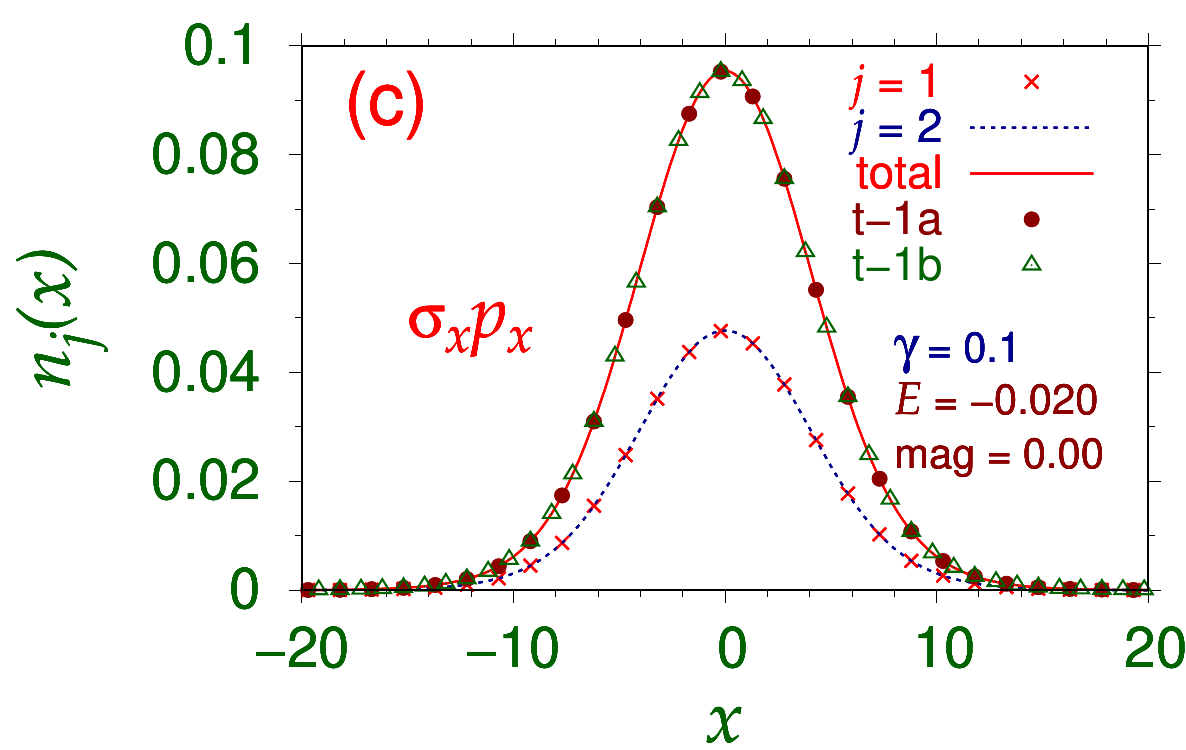}  
\includegraphics[width= .49\linewidth]{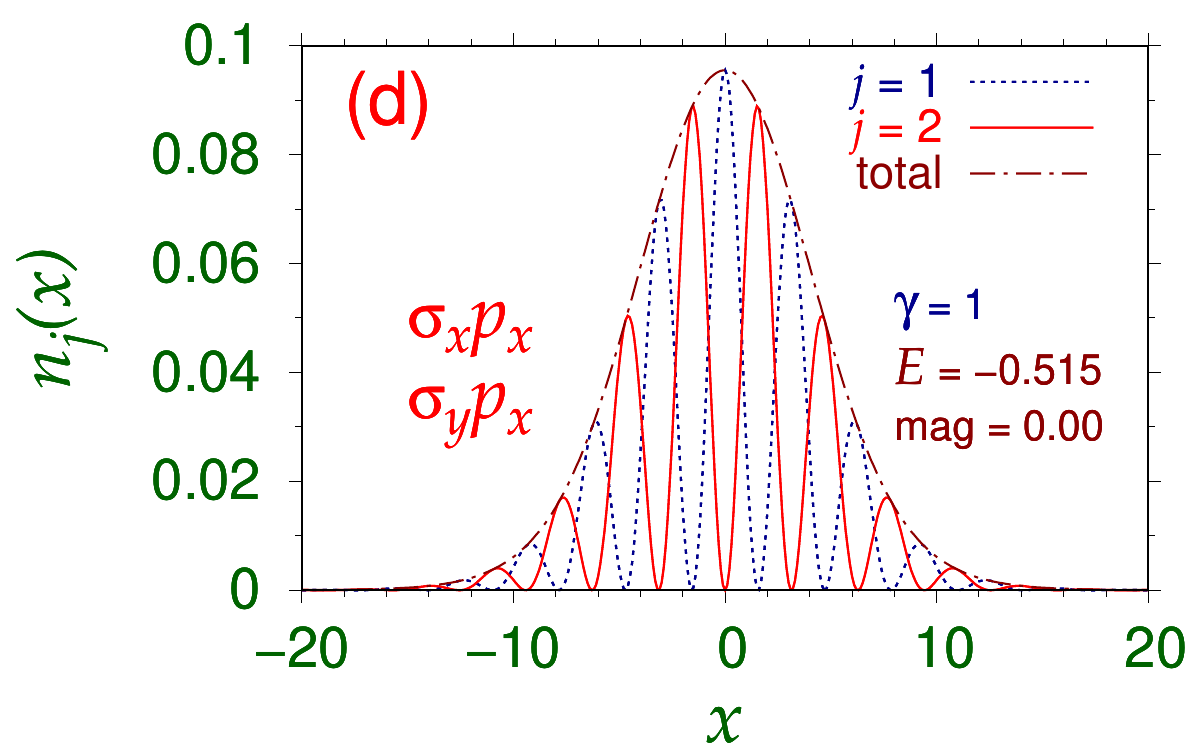} 
\includegraphics[width= .49\linewidth]{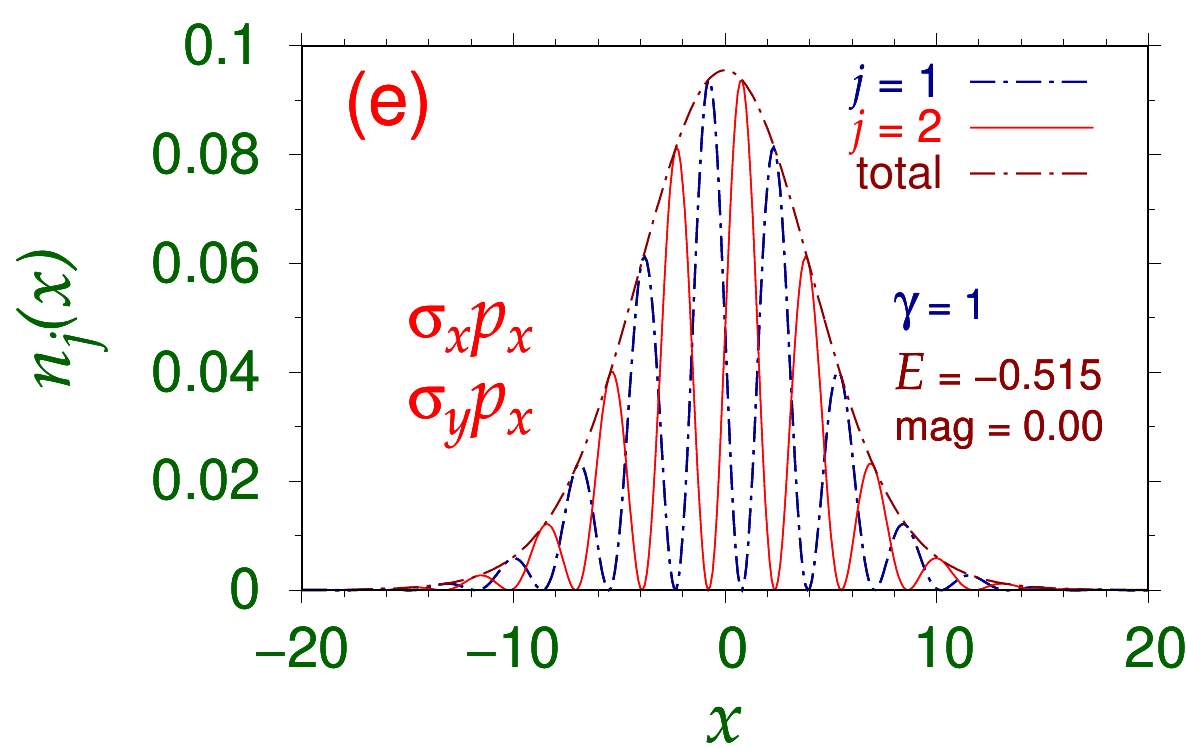}  
\includegraphics[width= .49\linewidth]{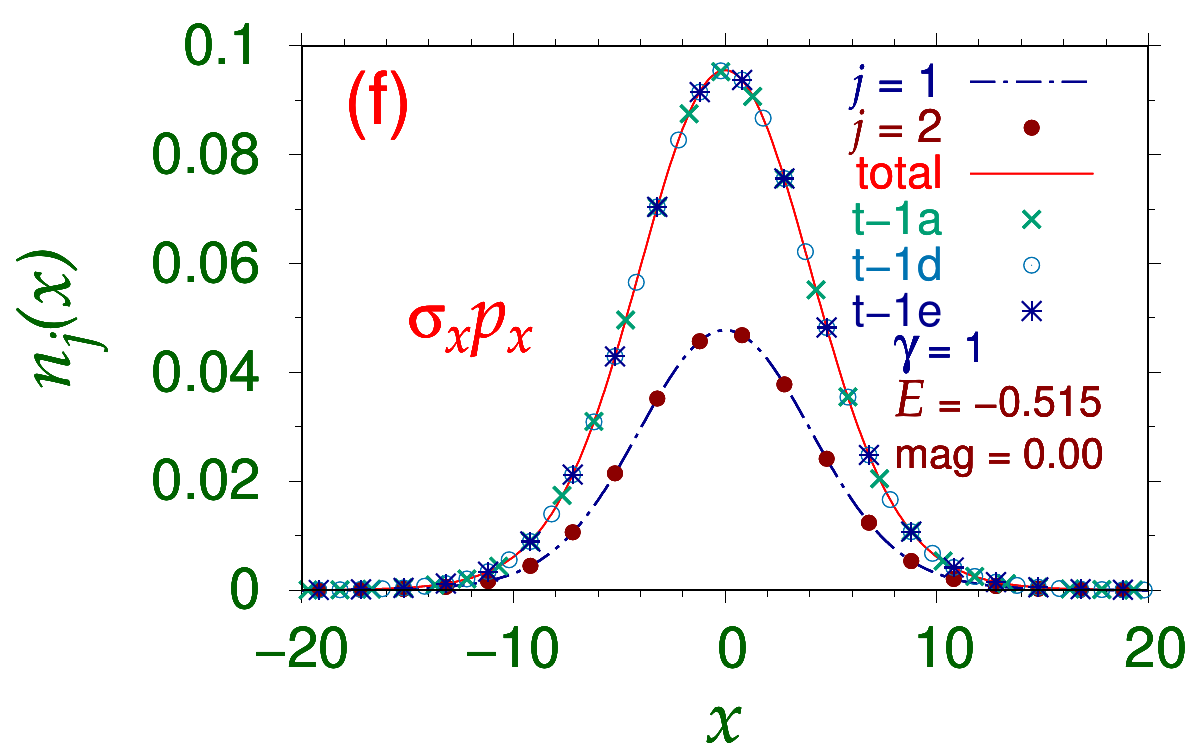}

\caption{(Color online) Density of components $j=1,2$  and total density of a quasi-1D  SO-coupled pseudo spin-half dipolar  (a) dark-bright,  (b) a phase-separated bright-bright, and (c) a bright-bright soliton for SO-coupling strength $\gamma =0.1$. In (c) we also show the total density of the solitons in plots (a)  (t-1a) and  (b)  (t-1b).
 The plots (d)-(f) display  a dark-bright,  a phase-separated     bright-bright, and a bright-bright   soliton,  for $\gamma =1$, with the first two having a spatially-periodic modulation in density. The energy ($E$), magnetization  (mag), and $\gamma$
 are given in the inset of all plots of this paper.
 The parameters are $c_0=c_2=0.5,  d=2$.   All quantities in this and following figures are
dimensionless and the total density is always normalized to unity, viz. Eq. (\ref{noma}).   }
\label{fig1}

\end{figure}

   Without losing generality, first we   consider  a dipolar  BEC with  a small value of the nonlinearity parameters  $c_0=c_2=0.5,  d=2$ in Eq. (\ref{EQ1}), corresponding to a repulsive intraspecies   and also  a  repulsive interspecies contact interaction.
   In this case, due to a moderate dipolar interaction, a soliton can be formed for a small value of the SO-coupling strength $\gamma$, which facilitates both theoretical and numerical investigations. For a small    $\gamma$ ($\gamma = 0.1$) we  find the following three types of  quasi-1D solitons in a self-repulsive dipolar BEC: 
a   dark-bright soliton,  an overlapping and a phase-separated  bright-bright soliton. 
 In Fig. \ref{fig1}  we present the component and total density of a quasi-1D SO-coupled  pseudo  spin-half  dipolar (a)  dark-bright, (b) a phase-separated bright-bright,  and (c) a bright-bright soliton for  SO-coupling strength $\gamma =0.1$ and SO couplings  $\gamma \sigma_x p_x$  and  $ \gamma \sigma_y p_x$.  The bright (dark) soliton component has a maximum (minimum of zero) of density at the center.  However,  the wave function in the presence of SO coupling is complex and  in all dark soliton components illustrated in  this paper,  at the central zero of density, there is a phase jump of $\pi$, as in any dark soliton \cite{dsx}.
The bright-bright soliton of Fig. \ref{fig1}(c) is  possible only for SO-coupling $\gamma \sigma_x p_x$.
These three solitons are quasi-degenerate with energy $E=-0.020$.  We use the term quasi-degenerate, as the  degeneracy is established only numerically. Also, we verified that the degeneracy is removed for a slightly larger value of the  nonlinearities. The total density $n$  of the solitons in Figs. \ref{fig1}(a)-(c) are equal as can be seen in Fig. \ref{fig1}(c), where we also plotted the total density $n$  of the solitons of  Figs. \ref{fig1}(a)-(b).    {  As we have different types of solitons for different (intraspecies and interspecies) interaction parameters,  these results are summarized in a tabular form in Table I, clearly
indicating which types of solitons emerge in the different parameter
regimes for both pseudo spin-half and spin-one dipolar BEC.

  \begin{table}[h!]
  {
    \centering
    \caption{Different types of soliton for dipolar spin-half and spin-one BECs:
    dark-bright (d-b), bright-bright (b-b), phase-separated bright-bright (ph-sep b-b), 
     dark-bright-dark (d-b-d), bright-dark-bright (b-d-b), bright-bright-bright (b-b-b).  
     The qualifier ``(m)'' denotes  a spatially periodic modulation in density of all components. The dipolar interaction parameter in all cases is $d=2$.
    }
    \label{tab:example_table}
    \begin{tabular}{|c|c|c|c|}
        \hline
       Spin & $\gamma$ &   $c_0, c_2$ & Type of soliton  \\
        \hline
     & & $c_0=0.5, c_2=0.5$&  d-b, ph-sep b-b, b-b \\
     1/2 & 0.1 & $c_0=-0.5, c_2=0.5$ & d-b, b-b \\
      &  & $c_0=0.5, c_2=-0.5$ & b-b \\
        \hline
    &  & $c_0=0.5, c_2=0.5$&  d-b (m), b-b (m), b-b\\
      1/2 & 1 & $c_0=-0.5, c_2=0.5$ &  d-b (m), b-b (m), b-b \\
      &  & $c_0=0.5, c_2=-0.5$ &  b-b \\
        \hline
            1 &0.1 &$c_0=0.5, c_2=0.5$ & ph-sep b-b-b, d-b-d, b-d-b \\
         & &$c_0=0.5, c_2=-0.1$ & b-b-b, ph-sep b-b-b \\
        \hline
                1 &1 &$c_0=0.5, c_2=0.5$ & d-b-d (m), b-d-b (m)\\
         & &$c_0=0.5, c_2=-0.1$ & b-b-b \\
        \hline
    \end{tabular}
    }
\end{table}

}

   As the SO-coupling 
   strength $\gamma$ is increased, the scenario of soliton formation changes; the three solitons of Figs. \ref{fig1}(a)-(c) evolve into the three solitons of Figs. \ref{fig1}(d)-(f), respectively.  For  $\gamma=1$ the linear density of the dark-bright soliton and  the phase-separated bright-bright soliton 
 develops a spatially-periodic modulation with the period  $\pi/\gamma$ as  illustrated in Figs. \ref{fig1}(d)-(e), respectively. In these cases   the total density has a smooth behavior without any spatial modulation.  The  density pattern of the overlapping bright-bright soliton remains the same  as    $\gamma $ is increased from   0.1 to  1, as illustrated in Figs. \ref{fig1}(c) and \ref{fig1}(f). The three solitons of Figs. \ref{fig1}(d)-(f) are quasi-degenerate with energy $E=-0.515$.   We could not find any other state  for these parameters. 
 Now we find that the total density of all the solitons in Figs. \ref{fig1}(a)-(f) are equal 
 as can be seen from Figs. \ref{fig1}(f), where we also plotted the total density of solitons of Figs. \ref{fig1}(a) and \ref{fig1}(d)-(e).  For a small $\gamma (\lessapprox 1)$, and small nonlinearities  
 $c_0,c_2,d,$ the degenerate ground states all have the same total density. 
   This could be possible if the energy is a function   of the total density. 
We verified that if $c_0$ and $c_2$ are different the degeneracy of the states for a fixed $\gamma$ will be removed and the solitons will not have the same total density.   
    As we are using $c_0=c_2$, using 
Eq. (\ref{sj})
the energy (\ref{ehalf})  can  be written as
\begin{align} \label{ehalf1}
 E &=\textstyle \frac{1}{2} \int_{-\infty}^\infty dx 
  \left[ \left\{ \sum_{j=1}^2|\partial _x \psi_j(x)|^2  -2{\mathrm i} \psi^T(x)  \gamma \partial_x \eta \psi(x) \right\}\right. \nonumber \\  &+\left.  \left\{c_0n^2(x)+d s(x)n(x)\right\}
  \right] \, , 
  \end{align}
  where
\begin{align} s(x)&=  \int \frac{d{ k}_x}{2\pi} e^{-{\mbox i} {k}_x x}
\widetilde n({k}_x)h_{1D}\left(\frac{k_x }{\sqrt {2 }}  \right) .
\end{align}
The energy (\ref{ehalf1}) has two parts. The first part in the first curly bracket
of this expression
 is the single-particle  linear Hamiltonian (\ref{abc1}), which contributes ${E_1}=-\gamma^2/2$ to energy as found in Sec. \ref{iii}.  The nonlinear interaction part in the second curly bracket of Eq. (\ref{ehalf1}) is a function of total density $n$, contributes an amount $E_2$,  independent of $\gamma$, to total energy $E
(=E_1+E_2)$. This provides a qualitative explanation of the  fact that all the  degenerate states of Fig. \ref{fig1} have the same total density.  From the quasi-degenerate states of Figs. \ref{fig1}(a)-(c), with $\gamma=0.1$ and $E=-0.020$, we can estimate  $E_2=E-E_1= -0.020 +\gamma^2/2=     -0.015$.   Using this $\gamma$-independent estimate for $E_2$ we can now analytically predict the energy $E$ of the three quasi-degenerate states of Figs. \ref{fig1}(d)-(f) for $\gamma=1$ to be $E=-\gamma^2/2 +E_2 = -0.515$  in agreement with the numerical 
result $E=-0.515$  of the solitons in Figs. \ref{fig1}(d)-(f).

   \begin{figure}[!t] 
\centering 
\includegraphics[width= .49\linewidth]{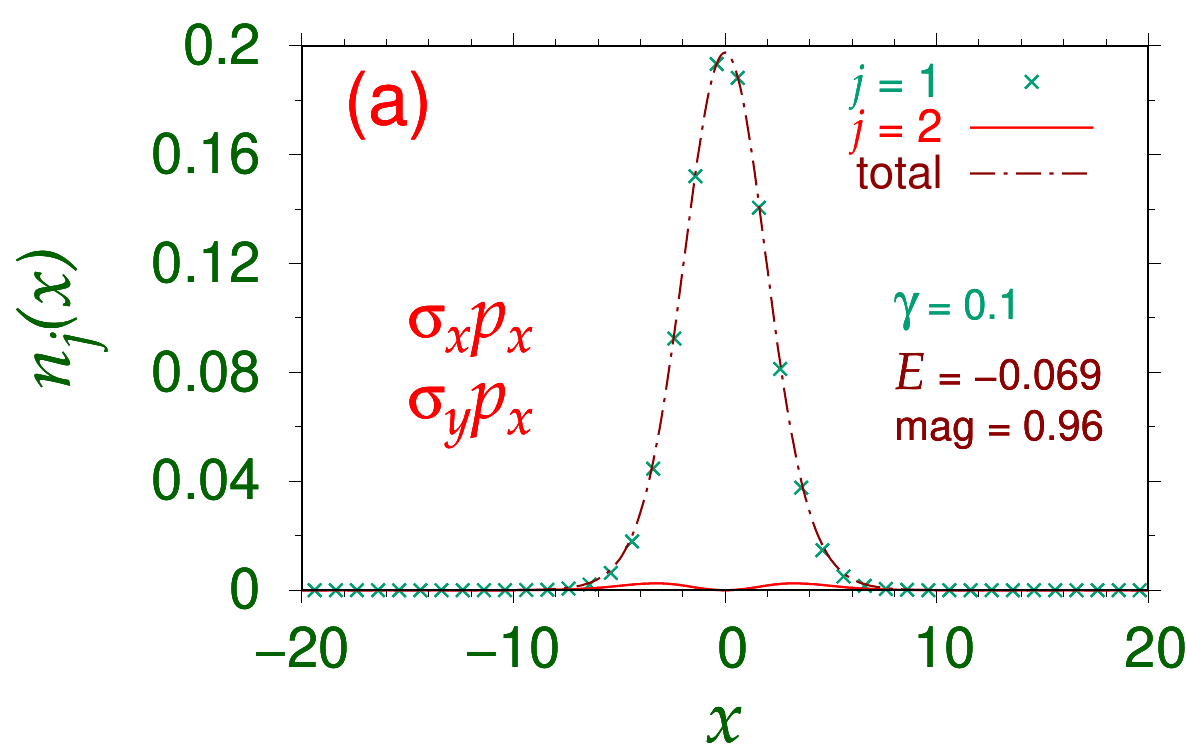}   
\includegraphics[width= .49\linewidth]{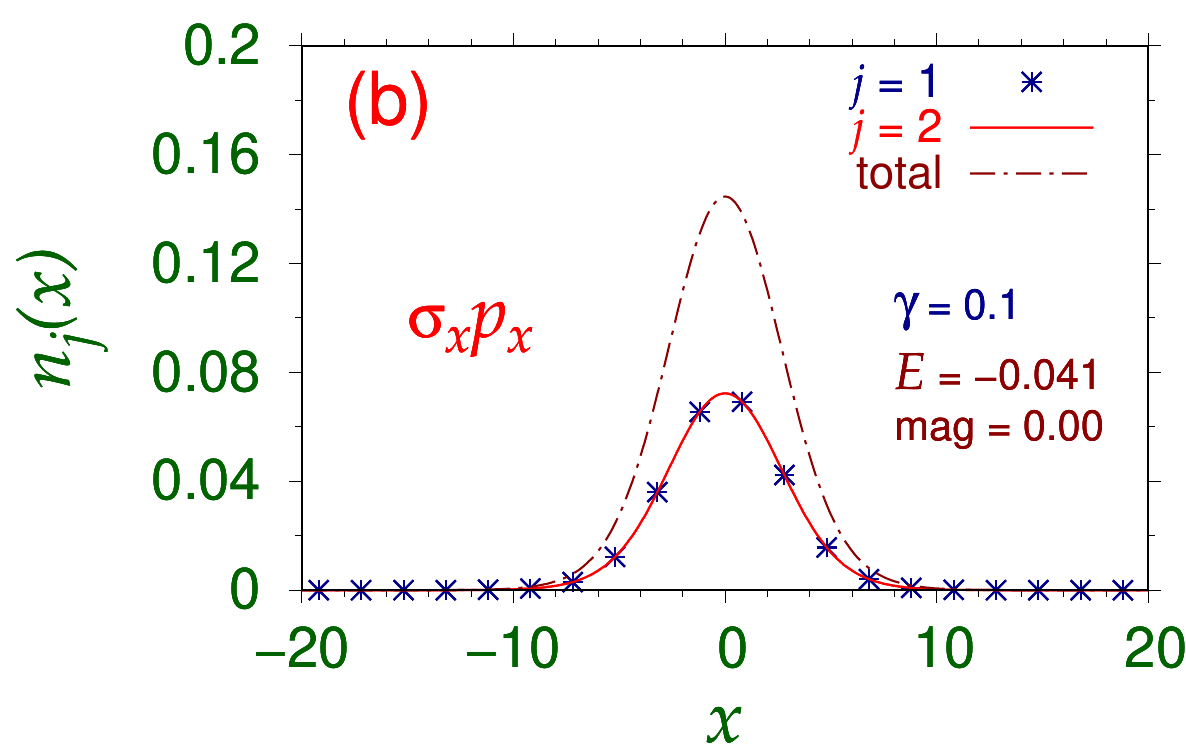}  
\includegraphics[width= .49\linewidth]{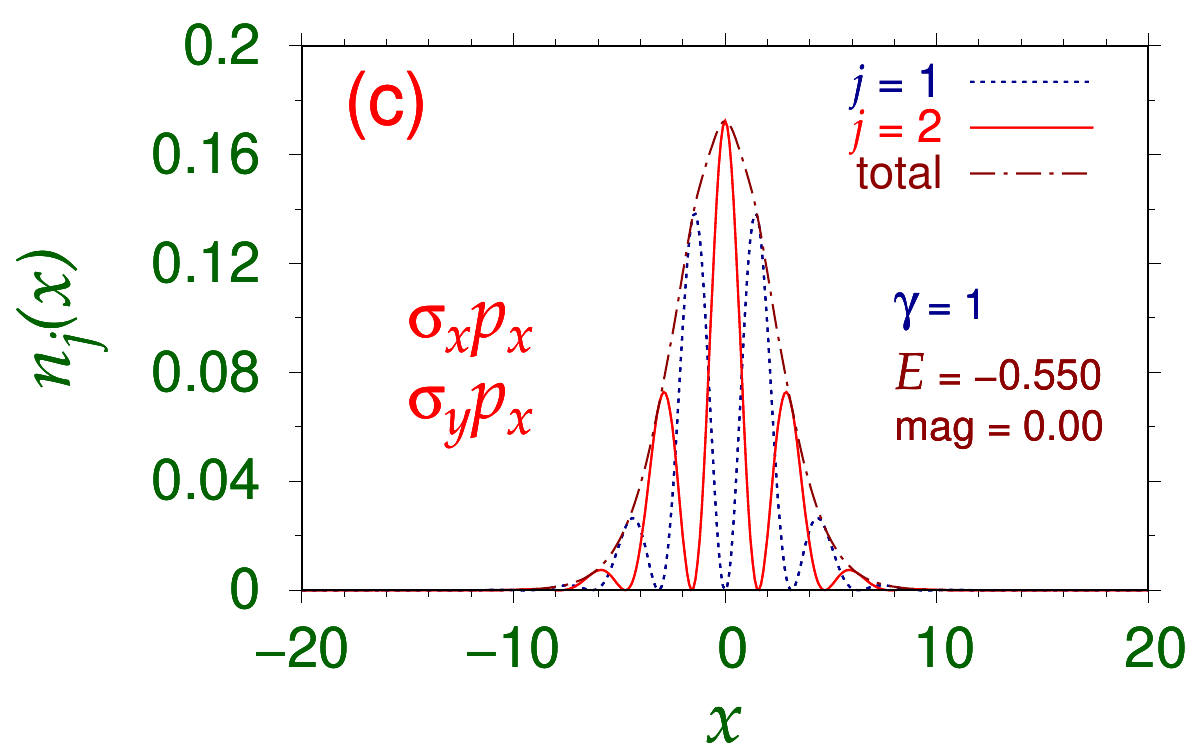}  
\includegraphics[width= .49\linewidth]{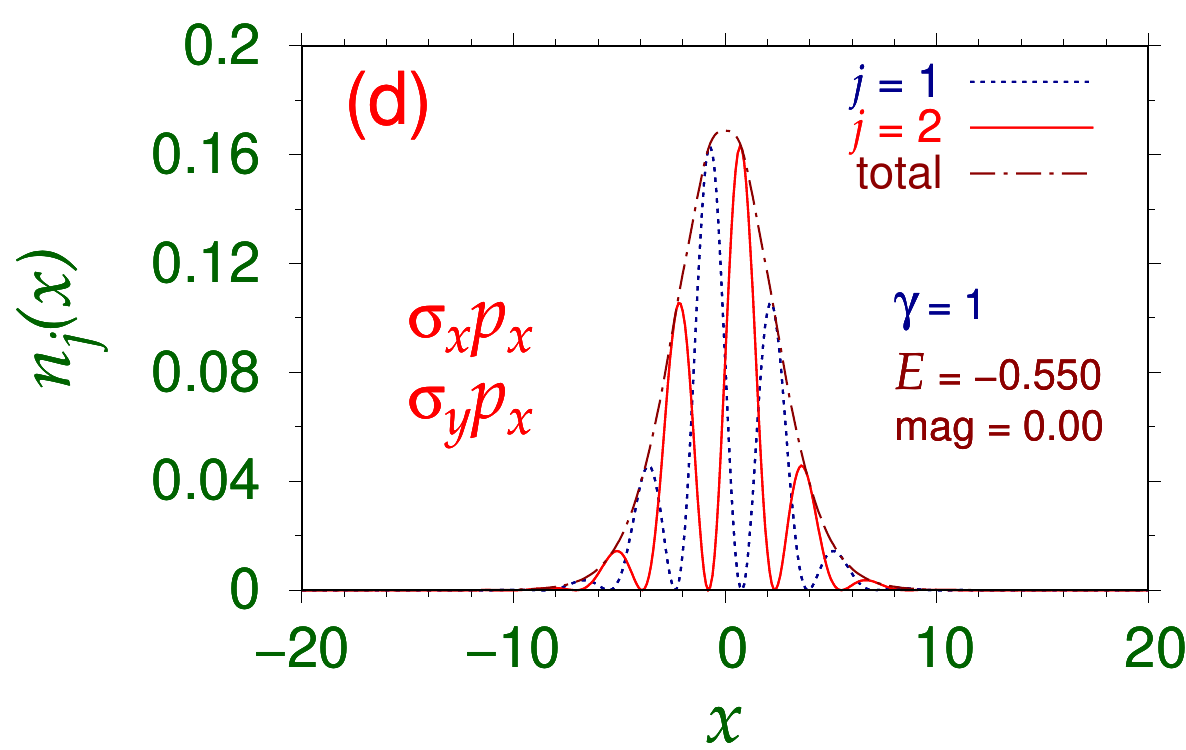} 
 \includegraphics[width= .49\linewidth]{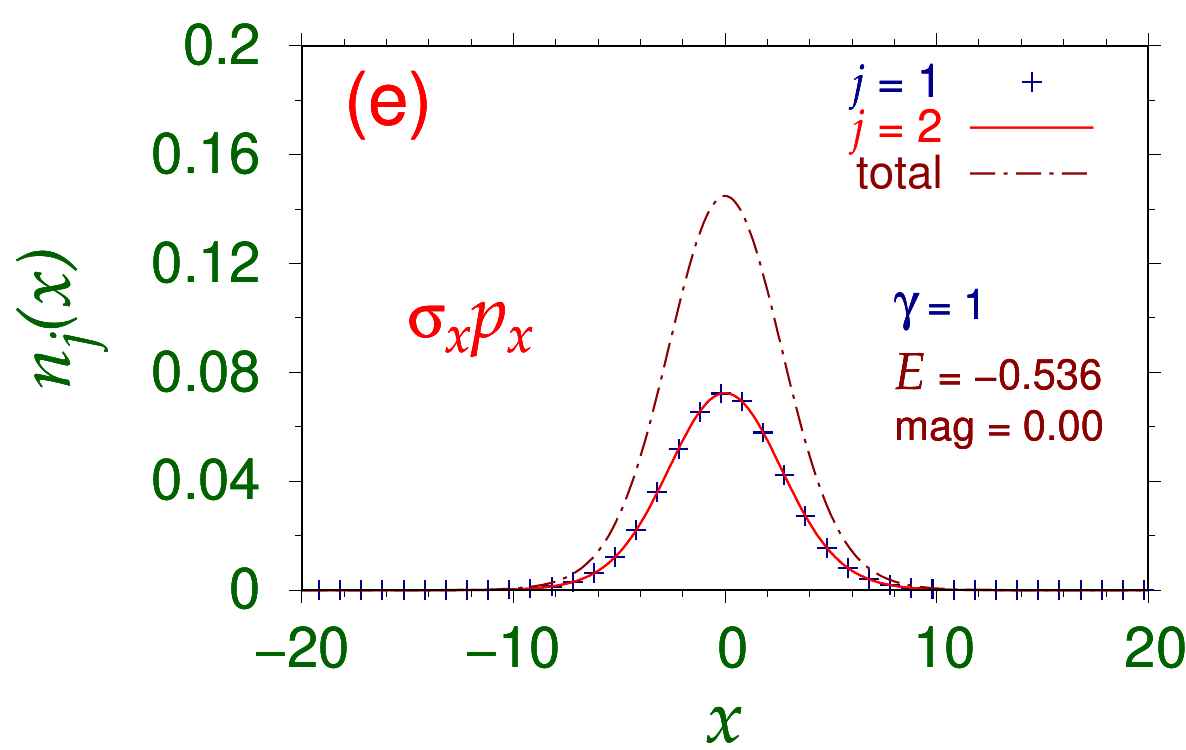}  
 \includegraphics[width= .49\linewidth]{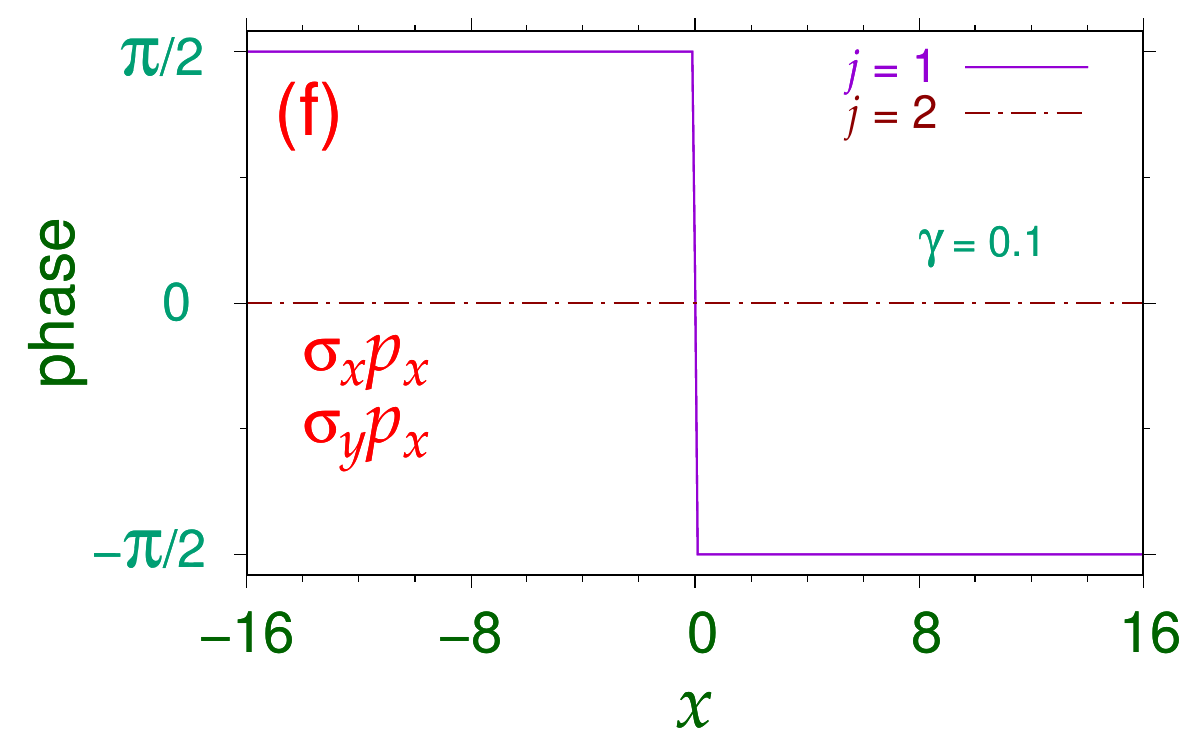}

\caption{Density of components $j=1,2$  and total density of a quasi-1D  SO-coupled pseudo spin-half dipolar  (a) dark-bright,   and (b)  bright-bright  soliton for SO-coupling strength $\gamma =0.1$.  The plots (c)-(e) display  a dark-bright,  a phase-separated     bright-bright, and a bright-bright   soliton, respectively  for $\gamma =1$.
The plot (f) presents the phase of the wave functions of the dark-bright soliton of plot (a). The parameters are $c_0=-0.5,  c_{2}=0.5,  d=2$.  }

\label{fig2} 

\end{figure}

However, the above breakup of energy $E$ in   $\gamma$-dependent linear part $E_1$ and
a $\gamma$-independent 
 nonlinear part $E_2$, viz. Eq. (40), is always possible,  for the small
values of the nonlinearities used in this paper, where, in general, $E_2$ will be a function of the component densities. If, for two states, the component densities are equal, these states should be degenerate if $\gamma$ is the same.  However, if    $\gamma$ of the two states is different, the energy of the two states are related by a simple algebraic relation
\begin{equation}\label{scale}
E^{(2)}=E^{(1)}-\frac{\gamma_2^2}{2}+\frac{\gamma_1^2}{2},
\end{equation}
where $E^{(2)}$ and $E^{(1)}$  are the energies of to states with equal component densities for $\gamma =\gamma_2$ and $\gamma_1$, respectively. The relation (\ref{scale})  is valid whenever the energy can be broken into two parts, and not only in the case of a  pseudo spin-half BEC.

   \begin{figure}[!t] 
\centering 
\includegraphics[width= .49\linewidth]{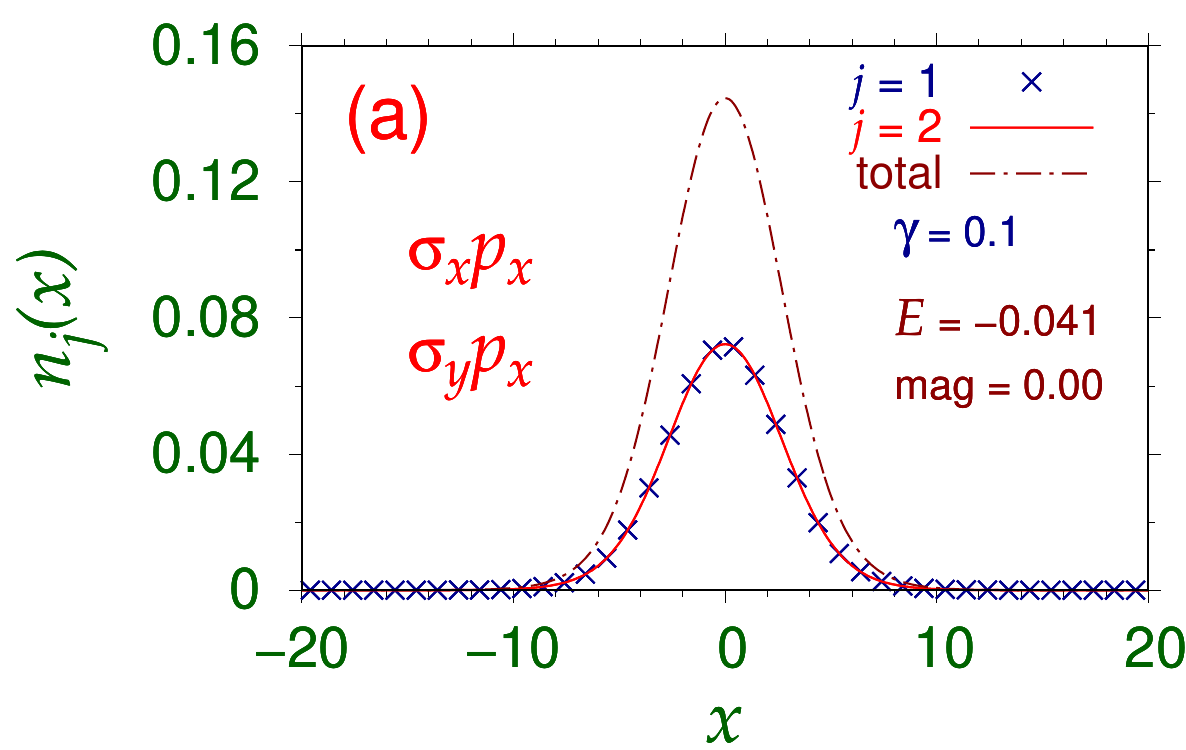}  
\includegraphics[width= .49\linewidth]{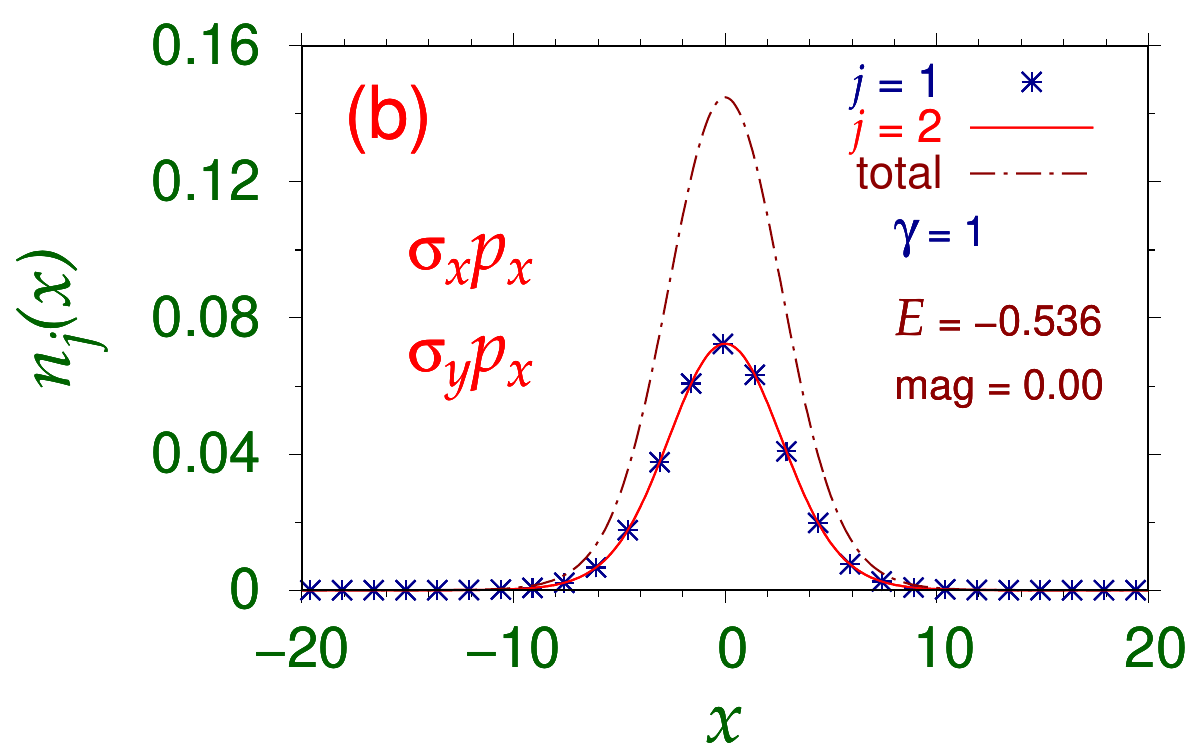}

\caption{Density of components $j=1,2$  and total density of a quasi-1D  SO-coupled pseudo spin-half dipolar    bright-bright soliton for  (a)  $\gamma =0.1$ and (b) $\gamma=1$.  
 The parameters are $c_0=0.5,  c_{2}=-0.5,  d=2$.    } 
\label{fig3}

\end{figure}

  To study the case of an attractive intraspecies interaction and a repulsive interspecies interaction we consider the parameters  $c_0=-0.5, c_{2}  =0.5,d=2.$ Although,
the intraspecies interaction is attractive, in the absence 
of a dipolar interaction ($d = 0$), and an SO coupling,
there is no binary soliton formation in this case for  the present set of parameters. However, there could be a separate (uncoupled) single-component soliton(s).  
  For $\gamma=0.1$  we have the  same 
  dark-bright  and bright-bright solitons, viz. Figs. \ref{fig1}(a) and \ref{fig1}(c),  as illustrated in Figs. \ref{fig2}(a)-(b), in increasing order of energy.  The soliton in Fig. \ref{fig2}(b) is in an excited  metastable  state.
  As $c_0 \ne c_2$ in this case, the total densities and energies of the solitons in Figs. \ref{fig2}(a)-(b) are different.
 In the case of the dark-bright soliton of Fig. \ref{fig2}(a),  the magnetization   is almost unity  (${|\cal M|}=0.97 \approx 1 $)  indicating a very small number of atoms in the dark soliton component.  However,   the phase-separated bright-bright soliton of the type presented in Fig. \ref{fig1}(b)  is not possible in this case for $\gamma =0.1$. For  $\gamma=1$  we have (the same three types of solitons as presented in Fig. \ref{fig1}): the  dark-bright, the phase-separated bright-bright   and  the overlapping bright-bright  solitons, as illustrated in Figs. \ref{fig2}(c)-(e), respectively.
   Again, the  dark-bright and  phase-separated bright-bright 
 solitons shown in Figs. \ref{fig2}(c)-(d) are quasi-degenerate ground states  and have a  spatially-periodic  modulation in density, whereas the the overlapping bright-bright  soliton of Fig. \ref{fig2}(e) is an excited  metastable state. The bright-bright solitons of Figs. \ref{fig2}(b) and  \ref{fig2}(e) are possible only for SO coupling $\gamma \sigma_x p_x$ and not for $\gamma \sigma_y p_x$.
 In this case the component and total densities of the   solitons in Figs. \ref{fig2}(b) and \ref{fig2}(e) for $\gamma_1 =0.1$ and $\gamma_2=1$, respectively, are identical and the two energies  $E_1=-0.041$ and $E_2=-0.536$ satisfy the relation (\ref{scale}).
 In Fig. \ref{fig2}(f) we plot the phase of the wave function of the two components of the dark-bright soliton presented in Fig. \ref{fig2}(a). As expected \cite{dsx}, the phase of the dark component shows a jump of $\pi$ at the origin while that of the bright component is zero. In fact, there is such a jump whenever there is a zero in the wave function leading to many such jumps in the phase, for example, in the case of the soliton in Fig. \ref{fig2}(d).

 Finally, we consider the case of a repulsive intraspecies interaction and an attractive interspecies interaction, employing the parameters $c_0=0.5, c_{2}=-0.5$ and $d=2$. For an SO-coupling strength $\gamma=0$  and dipolar interaction $d=0$, in this case there is no binary soliton, confirming net repulsion in the model.  For both $\gamma=0.1$ and 1, there is only a single state $-$ the  bright-bright  soliton illustrated in Figs. \ref{fig3}(a) and (b), respectively. The dark-bright soliton and the phase-separated bright-bright soliton are  not possible here,  as the imaginary-time propagation converges to a magnetization  $|\cal M|$ unity or no atoms in the dark component. There is also no soliton in this case with spatially periodic stripe in density. The strong interspecies contact attraction and  the dipolar attraction sqeeze the solitons to the central region and do  not permit the formation of density modulation over a larger region of space.
  In this case the component and total densities of the solitons in Figs. \ref{fig3}(a) and \ref{fig3}(b) for $\gamma_1 =0.1$ and $\gamma_2=1$, respectively, are identical and the two energies  $E_1=-0.041$ and $E_2=-0.536$
  satisfy the relation (\ref{scale}).

\subsection{ Quasi-1D SO-coupled    spin-one self-repulsive  dipolar BEC soliton}
 
\label{b}

We now consider the scenario of soliton formation in a quasi-1D self-repulsive SO-coupled dipolar spin-one BEC. In this case there are two contact interaction parameters $c_0$ and $c_2$. First, we consider an anti-ferromagnetic  BEC originating from a positive $c_2 >0$ \cite{ku}. To study the case of an anti-ferromagnetic  BEC, we consider the parameters 
$c_0=c_2=0.5, d=2.$  In this case also, for $d=\gamma =0$, there is no soliton formation.
We  find three types of solitons in this case: partially phase-separated  bright-bright-bright,  dark-bright-dark, and bright-dark-bright solitons  as displayed in Figs. \ref{fig4}(a)-(c), in increasing order of energy.  The solitons in Figs. \ref{fig4}(b)-(c) are excited metastable states.  The partially phase-separated bright-bright-bright soliton can only be obtained with the SO coupling $\gamma \Sigma_x p_x$, all other solitons in this case can be obtained with both types of SO coupling 
 $\gamma \Sigma_x p_x$ and  $\gamma \Sigma_y p_x$.  
  For a large $\gamma =1$, there are only following two quasi-degenerate solitons with a spatially-periodic modulation in density: dark-bright-dark and bright-dark-bright solitons illustrated in Figs. \ref{fig4}(d)-(e), respectively.   In Fig. \ref{fig4}(f) we plot the phase of the wave function of the three components of the dark-bright-dark soliton presented in Fig. \ref{fig4}(a). As expected \cite{dsx} the phase of the dark components shows a jump of $\pi$ at the origin while that of the bright component is zero.

 \begin{figure}[!t] 
\centering 
\includegraphics[width= .461\linewidth]{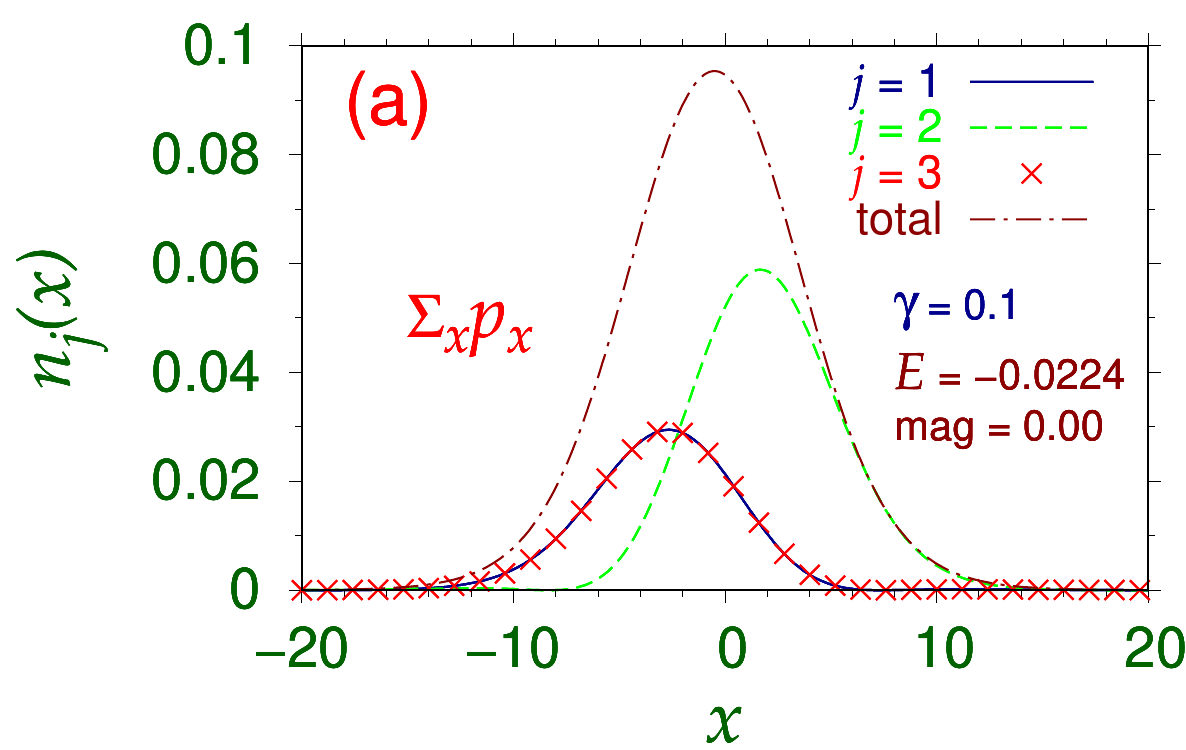}  
\includegraphics[width= .461\linewidth]{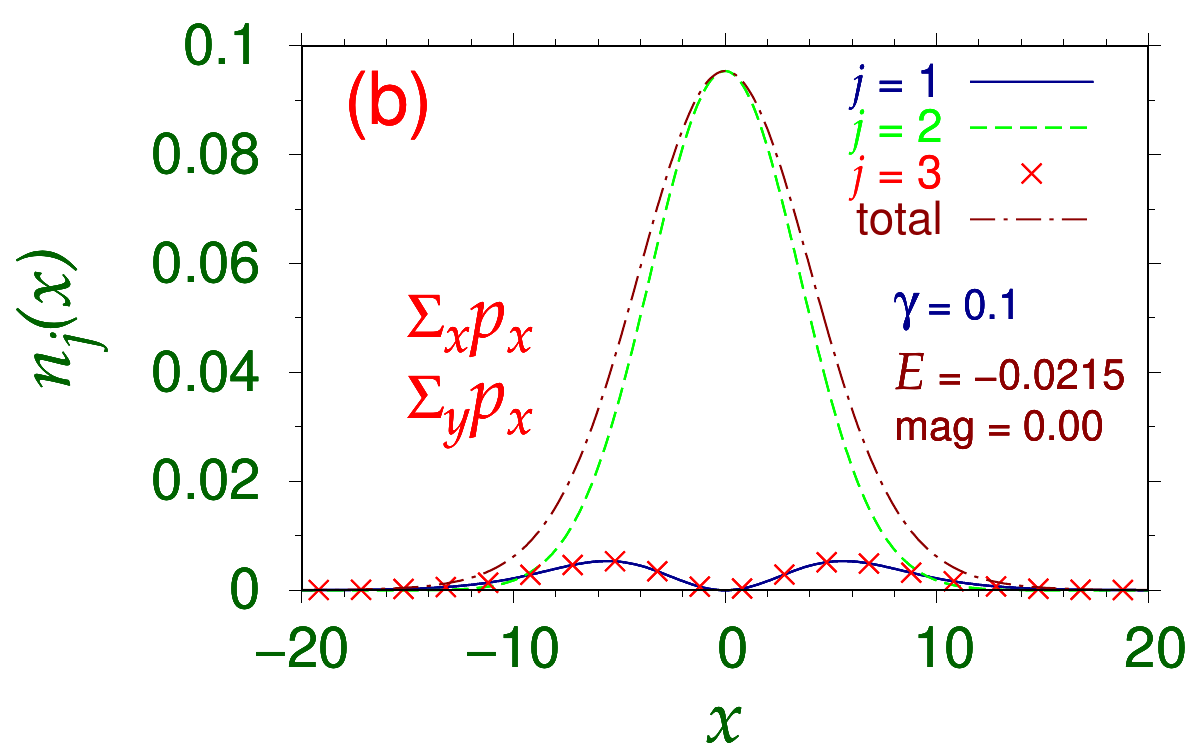}  
\includegraphics[width= .461\linewidth]{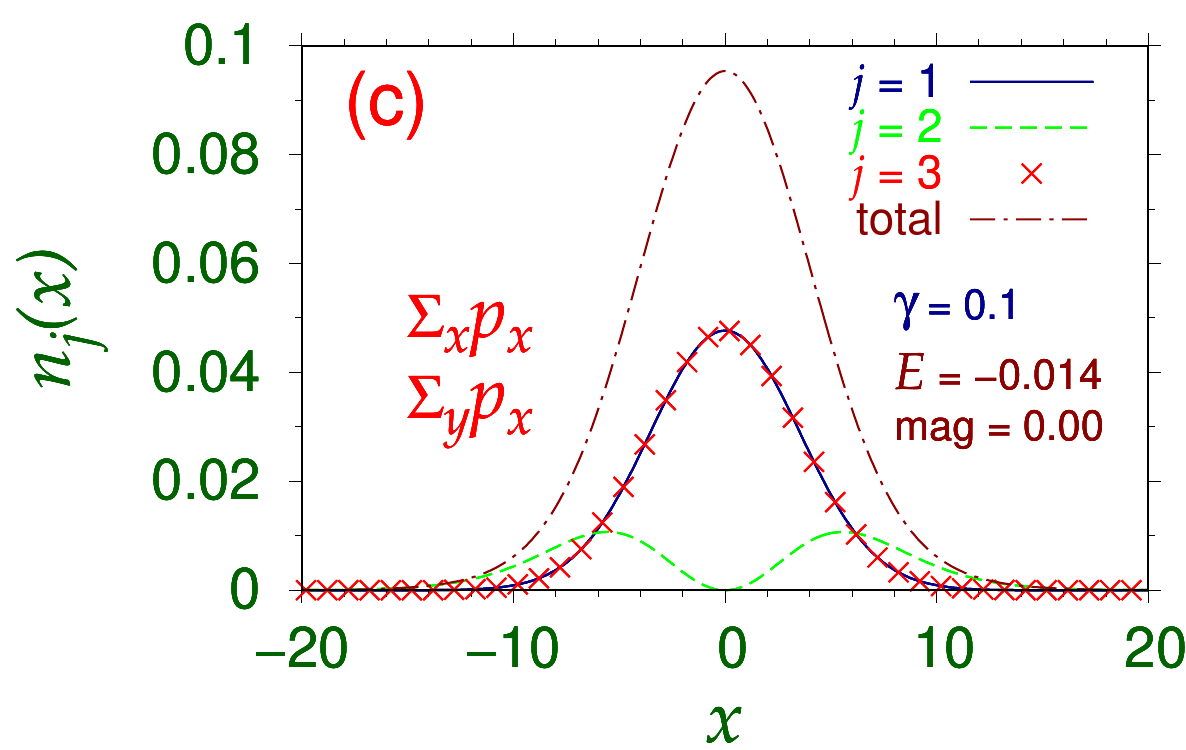}  
\includegraphics[width= .461\linewidth]{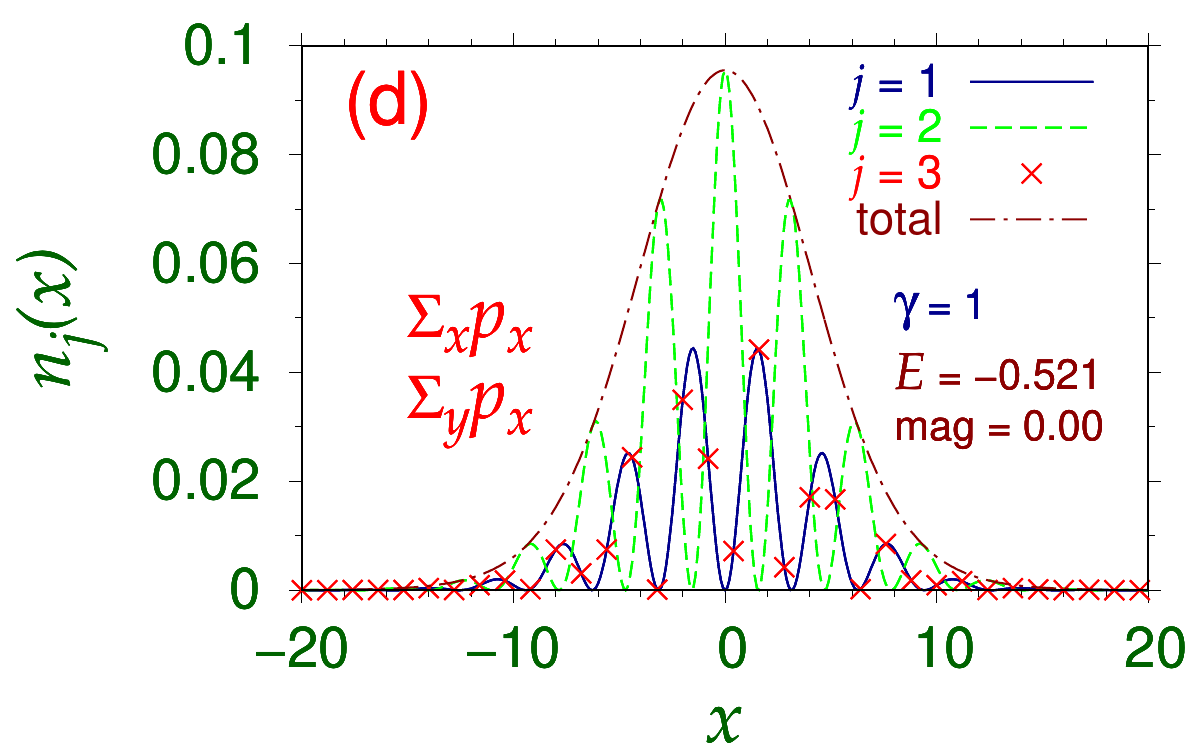} 
\includegraphics[width= .461\linewidth]{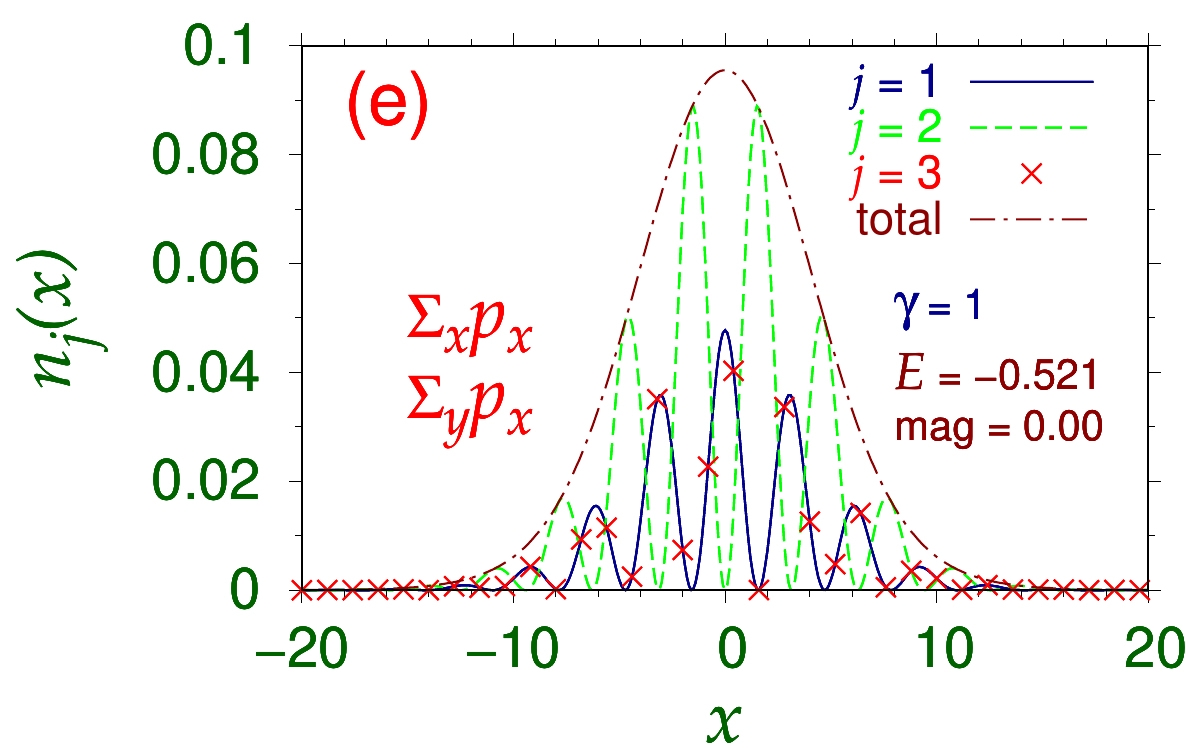}  
\includegraphics[width= .461\linewidth]{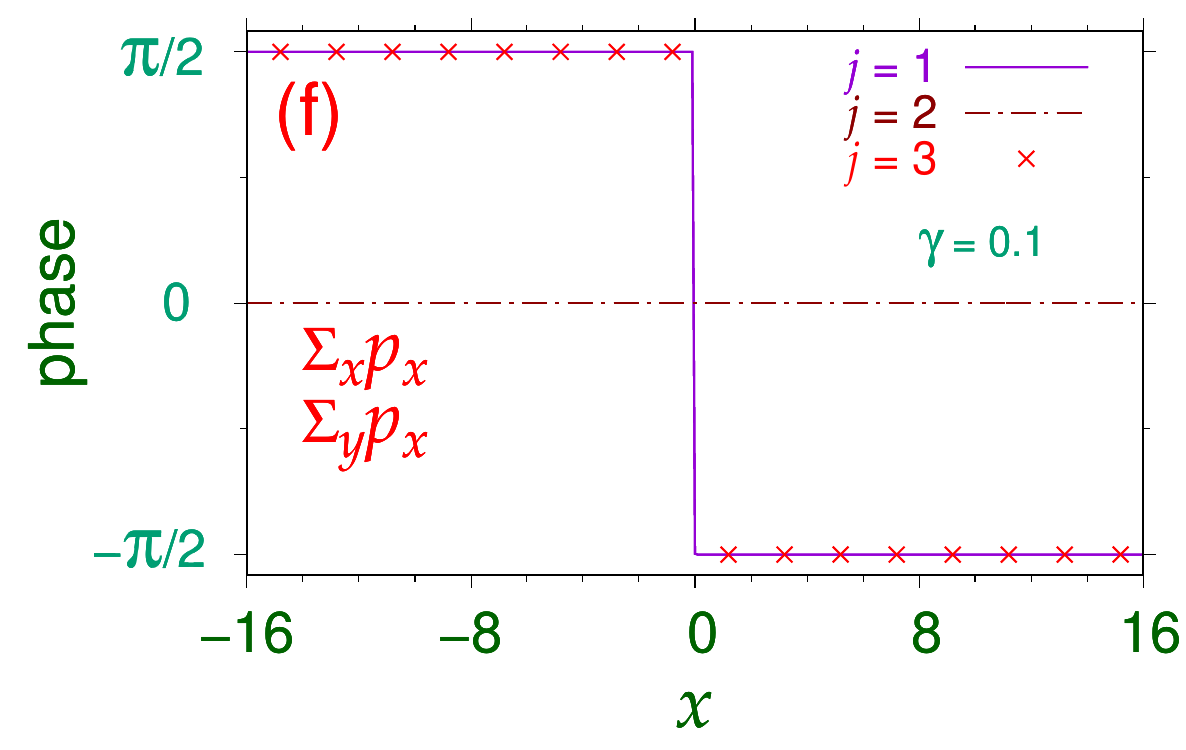}

\caption{ Density of components $j=1,2,3$  and total density of a quasi-1D  SO-coupled spin-one dipolar  (a) dark-bright-dark,  (b)  partially phase-separated  bright-bright-bright and (c) bright-dark-bright  solitons for SO-coupling strength $\gamma =0.1$ for an anti-ferromagnetic 
spin-one BEC.  The plots (d)-(e) display    a dark-bright-dark and a bright-dark-bright   soliton, both with a spatially-periodic modulation in density,  respectively  for $\gamma =1$. The plot (f) depicts the phase of the component wave functions of the soliton of plot (a).  The parameters are $c_0=0.5, c_2=0.5, d=2$. }

\label{fig4} 

\end{figure}

 Next we study the formation of quasi-1D solitons in a ferromagnetic SO-coupled spin-one dipolar BEC employing  the parameters $c_0=0.5, c_2=-0.1,d=2$. In this case also, for $d=\gamma =0$, there is no soliton formation.
 For $\gamma=0.1$,  one can have only two types of  distinct solitons, as illustrated  in Fig. \ref{fig5},  (a) a bright-bright-bright soliton and (b) a  phase-separated bright-bright-bright soliton.  This phase-separated  bright-bright-bright soliton was obtained by introducing a relative phase of $-\pi$ between components $j=1$ and 3 in the initial state to generate a repulsion between these two components.  Because of this additional repulsion, the   phase-separated bright-bright-bright soliton of Fig. \ref{fig5}(b) appears as  an excited  metastable state.
 The phase-separated soliton of Fig. \ref{fig5}(b) is distinct from the partially phase-separated bright-bright-bright soliton in the anti-ferromagnetic case, viz. Fig. \ref{fig4}(a).
 For $\gamma=0.1$, the quasi-1D solitons in  the ferromagnetic BEC are  completely  distinct  from the quasi-1D solitons  in the anti-ferromagnetic case, viz. Fig. \ref{fig4}(a)-(c), with no dark soliton component in the  ferromagnetic case. 
 { For a  large SO-coupling strength ($\gamma=1$),  the  only possible soliton is of the bright-bright-bright type as shown in Figs.  \ref{fig5}(c).}
   In this case the component and total densities of the solitons in Figs. \ref{fig5}(a) and \ref{fig5}(e) for $\gamma_1 =0.1$ and $\gamma_2=1$, respectively, are identical and the two energies 
 $E_1=-0.026$ and $E_2=-0.521$  satisfy the relation (\ref{scale}).

 \subsection{Dynamical stability of the solitons}
 \label{c}
 
 {  The  solitons presented in Figs. \ref{fig1}-\ref{fig5}, for the same set of parameters,   are often  quasi-degenerate,  viz.  Figs.  \ref{fig1}(a)-(c); sometimes they are excited states, viz. Figs. \ref{fig2}(b),  \ref{fig2}(e), \ref{fig4}(b)-(c) and \ref{fig5}(b). This could make their experimental observation difficult, as these states  could transform to their degenerate counterpart or else they can decay to the ground state.  Imaginary-time propagation method usually finds the lowest-energy state  with a  definite symmetry property. For example, in a quasi-1D  single-component harmonically trapped BEC it can find the lowest-energy parity-symmetric state (ground state) as well as the lowest-energy parity-antisymmetric state (the first excited state), which is a dark soliton with a zero density  at the middle. 
The dark soliton  state, although obtained by imaginary-time propagation approach, is dynamically unstable, as it eventually decays to the parity-symmetric ground state.  This makes the experimental observation of dark  solitons very difficult.  The solitons can be observed if they are dynamically stable. 
 }

 \begin{figure}[!t] 
\centering 
\includegraphics[width= .49\linewidth]{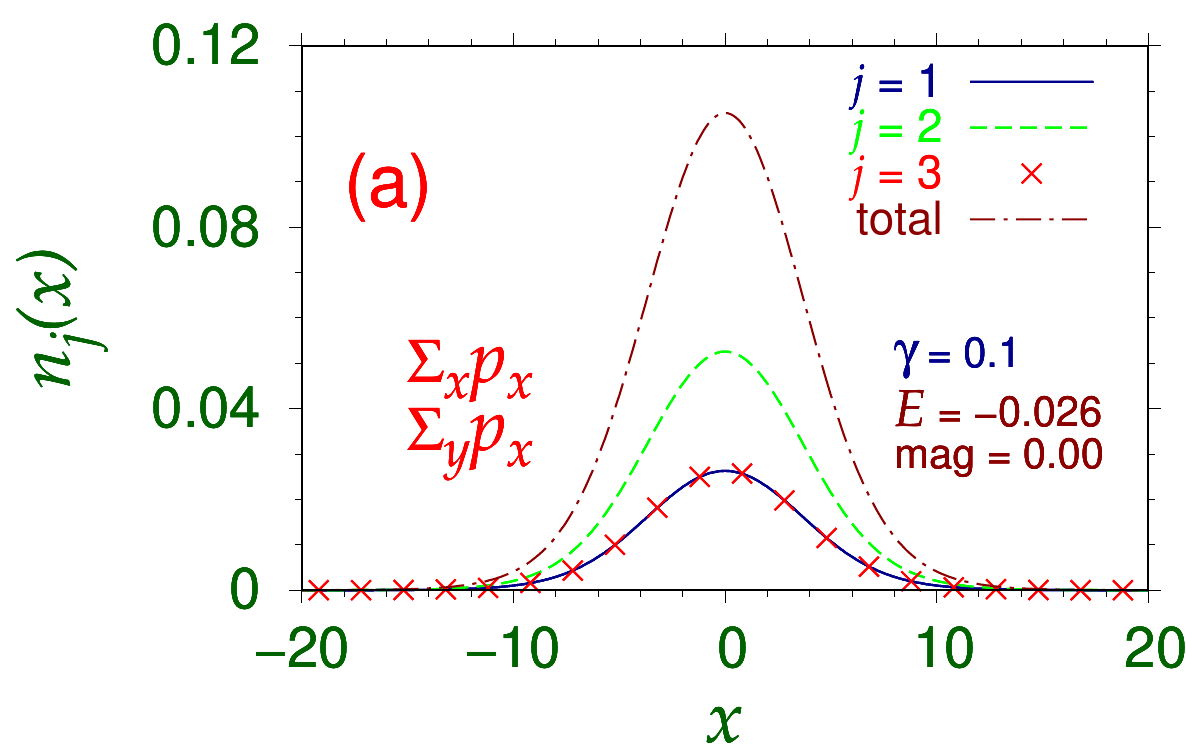}  
\includegraphics[width= .49\linewidth]{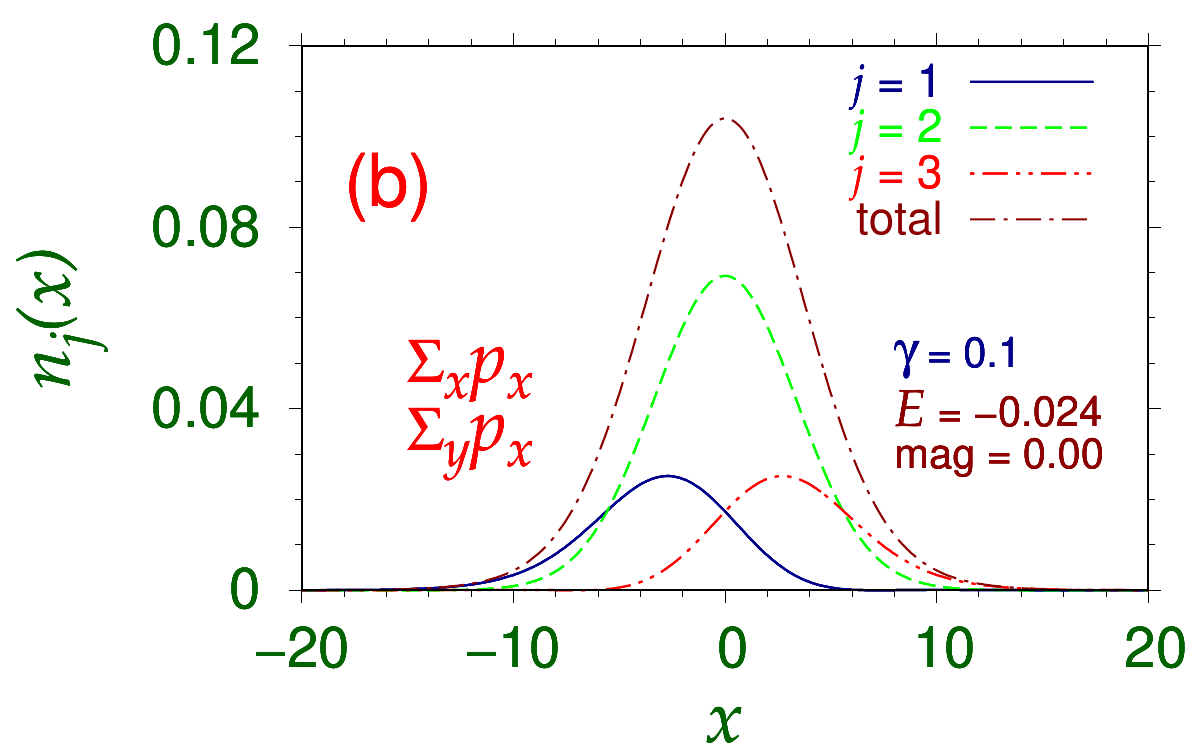}   
\includegraphics[width= .49\linewidth]{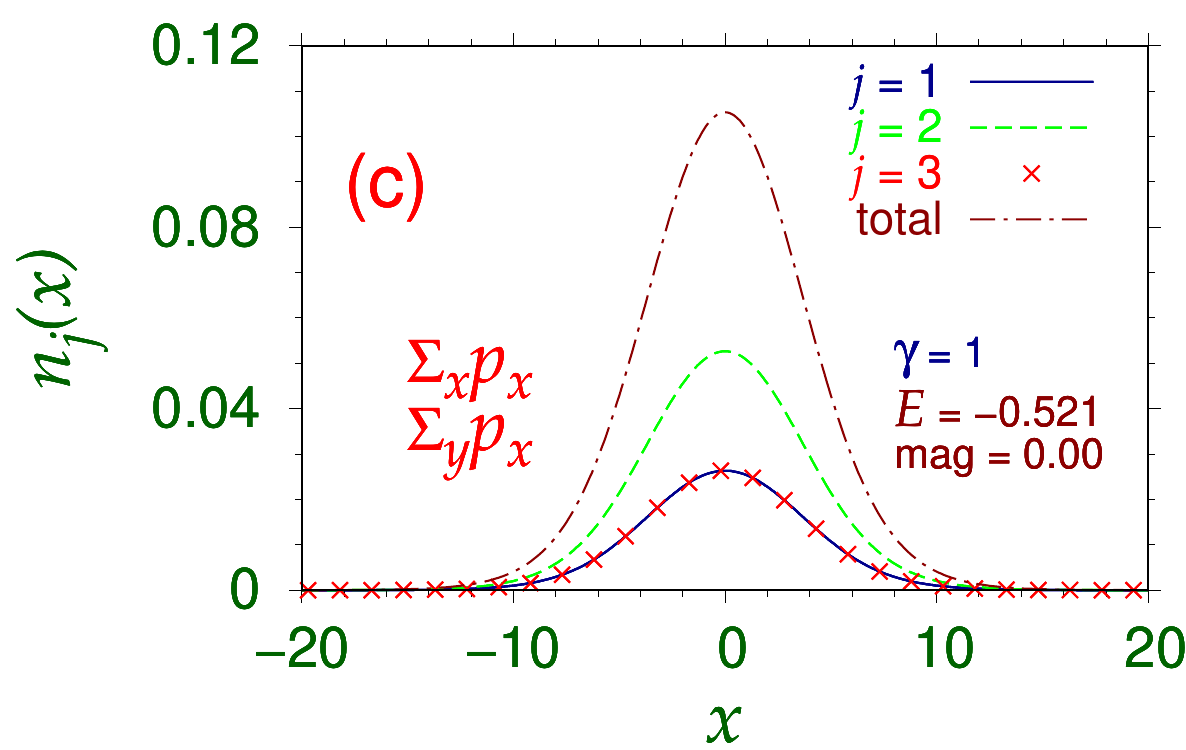}   

\caption{ Density of components $j=1,2,3$  and total density of a quasi-1D  SO-coupled spin-one dipolar  (a)  bright-bright-bright,  (b) a phase-separated bright-bright-bright   solitons for SO-coupling strength $\gamma =0.1$ for a ferromagnetic 
spin-one BEC.   {The plot (c) displays a bright-bright-bright    soliton for $\gamma =1$.}
  The parameters are $c_0=0.5, c_2=-0.1, d=2$. }
\label{fig5}

\end{figure}

 \begin{figure}[!t] 
\centering 
\includegraphics[width= .49\linewidth]{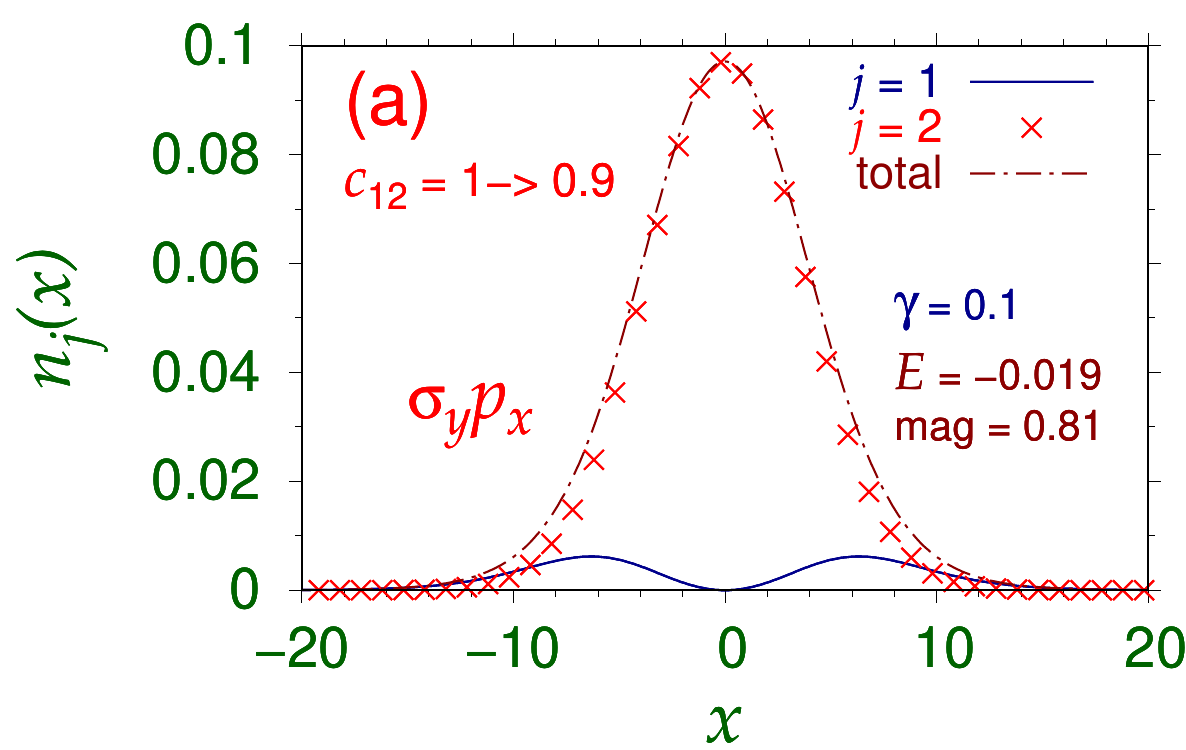}  
\includegraphics[width= .49\linewidth]{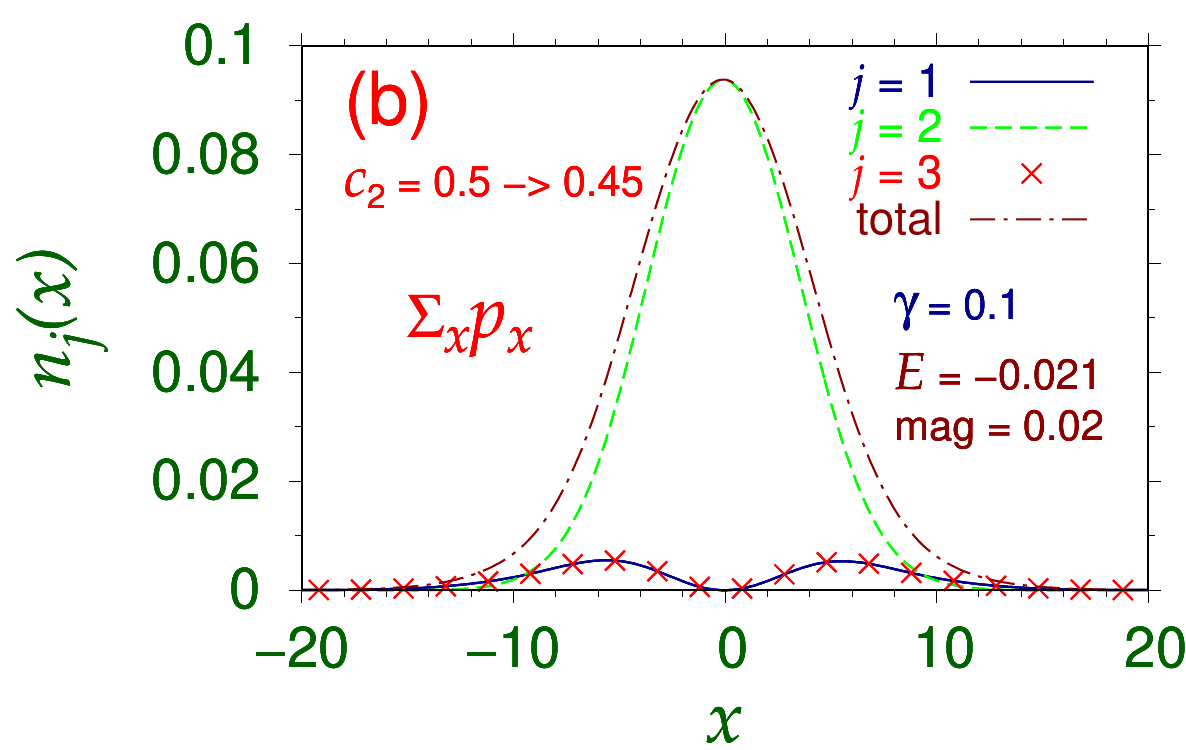}

\caption{ (a)Density of components $j=1,2$  and total density of the quasi-1D  SO-coupled pseudo spin-half dark-bright  dipolar soliton  of Fig. \ref{fig1}(a)  after  real-time propagation for  250 units of time, while the contact interaction parameter $c_{2}$  was changed from 1 to 0.9 at the start of the dynamics. 
(b) Density of components $j=1,2,3$  and total density of the quasi-1D  SO-coupled spin-one dark-bright-dark   dipolar soliton  of Fig. \ref{fig4}(a)  after  real-time propagation for  250 units of time, while the contact interaction parameter $c_{2}$  was changed from 0.5  to 0.45 at the start of the dynamics. 
  }

\label{fig6}

\end{figure}

 To study the dynamical stability of the present SO-coupled quasi-1D pseudo spin-half and spin-one dipolar solitons, we pay special attention to the soliton complexes with a dark soliton component, because dark solitons in a single-component BEC are unstable in general \cite{ds,ds1,ds3,ds4}.  We find that all the solitons studied in this paper are dynamically stable.
 We present results of stability of the pseudo spin-half dark-bright soliton  depicted in Fig. \ref{fig1}(a) and the spin-one dark-bright-dark soliton illustrated in Fig. \ref{fig4}(a), both with a dark soliton component. First, we  consider the converged imaginary-time wave function of the pseudo spin-half SO-coupled dipolar dark-bright soliton for $\gamma=0.1$, viz.  Fig. \ref{fig1}(a), and perform real-time simulation  after changing the nonlinearity $c_{2}$ from 1 to  0.9 at time $t=0$  using the imaginary-time wave function as the initial state.  The resultant real-time  density at time $t=250$ units of time is displayed in Fig. \ref{fig6}(a), which compares well with the imaginary-time density of Fig. \ref{fig1}(a), demonstrating the dynamical stability of the soliton. Although,
 the final  energy $E=-0.19$ after real-time propagation in Fig. \ref{fig6}(a)  is quite close to the imaginary-time energy $E=-0.20$, the final magnetization ${\cal M} =-0.81$ after real-time propagation is a bit different from the imaginary-time magnetization   
  ${\cal M} =-0.69$. This difference in magnetization is due to a decrease in interspecies 
  repulsion in real-time propagation.

 To establish the dynamical stability of the spin-one dark-bright-dark SO-coupled dipolar soliton of Fig. \ref{fig4}(a), we use the  converged imaginary-time wave function of this soliton as the initial state in a real-time propagation after changing the nonlinearity parameter $c_2$ from 0.5 to 0.45 at time $t=0$. The resultant real-time density at time $t=250$ units of time is displayed in Fig. \ref{fig6}(b), in good agreement of the imaginary-time density of Fig. \ref{fig4}(a), which demonstrates the dynamical stability.
  The final  energy $E=-0.21$ after real-time propagation in Fig. \ref{fig6}(b)  is quite close to the imaginary-time energy $E=-0.22$.

\section{Summary and Discussion}
 \label{d}

To search for a quasi-1D soliton  in a uniform   pseudo spin-half  and spin-one SO-coupled dipolar, self-repulsive  BEC, using the numerical solution of a mean-field model,  we identify   different types of solitons.   In the absence  of an  SO coupling and a dipolar interaction, the net prevailing interaction in all the systems, considered in this paper,  is repulsive and  no solitons can be formed, which justifies the term  self-repulsive. In the pseudo spin-half case, for  repulsive 
intraspecies and interspecies contact interactions, for a small SO-coupling strength ($\gamma =0.1$),  there are dark-bright, and phase-separated  and overlapping bright-bright  solitons, viz. Figs. \ref{fig1}(a)-(c). For an  attractive  intraspecies and a repulsive  interspecies contact interaction,
for a small $\gamma$,  we have only dark-bright and  bright-bright  solitons, viz. Figs. \ref{fig2}(a)-(b).  For a repulsive 
intraspecies and an attractive interspecies contact interaction,  for all  $\gamma$ ($=0.1,$ and $ 1$), only the overlapping bright-bright soliton of Fig. \ref{fig1}(c) is possible, viz. Fig. \ref{fig3}(a)-(b).
 For a large SO-coupling strength ($\gamma=1$), for repulsive  interspecies interaction and both attractive and repulsive intraspecies interactions,  the scenario of soliton formation is the same and we can  have  (A) a dark-bright soliton,  (B) a phase-separated bright-bright soliton,  and (C) an overlapping bright-bright soliton, viz. Figs. \ref{fig1}(d)-(f) and  Figs.  \ref{fig2}(c)-(e).  The first two of these solitons, e.g. (A) and (B), are quasi-degenerate  and also have a spatially-periodic modulation in density.  
 The  overlapping bright-bright  soliton of Fig. \ref{fig1}(f) is also degenerate with the other two solutions, whereas that of Fig. \ref{fig2}(e) is an  excited metastable state.
  
In the spin-one anti-ferromagnetic case  ($c_2>0$),   for a weak SO coupling ($\gamma=0.1$), there are three different types of solitons: a partially overlapping bright-bright-bright  soliton, viz. Fig. \ref{fig4}(a),  a dark-bright-dark soliton, viz. Fig. \ref{fig4}(b), and a  bright-dark-bright soliton, viz. Fig. \ref{fig4}(c). In the ferromagnetic case, for $c_2<0$, $\gamma =0.1$, we have completely different types of solitons: a bright-bright-bright soliton, viz. Fig. \ref{fig5}(a),  and  a phase-separated bright-bright-bright soliton, viz. Fig.    \ref{fig5}(b). For a large $\gamma =1$, in the anti-ferromagnetic case we have  only a bright-dark-bright and a dark-bright-dark quasi-degenerate solitons with a spatially-periodic stripe in density as shown in Figs. \ref{fig4}(d)-(e).   In the ferromagnetic case ($c_2<0$), for $\gamma =1$,  
{we only have } a bright-bright-bright soliton, viz. \ref{fig5}(c).

The spatially-periodic modulation in density of these solitons  \cite{14,baym1,2020,stripe}   is a    manifestation of supersolid formation \cite{s1,s2,s3,s4} in these systems. This supersolid formation was  observed experimentally in a pseudo spin-half SO-coupled BEC of $^{23}$Na atoms \cite{14}. Although,   a spatially-periodic modulation in density with a  period of $\pi/\gamma$  is clearly seen in the component densities in some cases,  there is no such modulation in the total density, consistent with the analytic consideration in Secs. \ref{iii} and \ref{iv}.

The dark soliton in a single-component BEC or in nonlinear optics, being  an excited state,  is  dynamically unstable. All solitons presented in this paper are dynamically stable, specially those  with a dark soliton component. We  demonstrated this for the pseudo spin-half dark-bright soliton of Fig. \ref{fig1}(a) and the spin-one bright-dark-bright soliton of Fig. \ref{fig4}(a)   by real-time propagation  over a long time interval after introducing a perturbation  at the start of the dynamics, viz. Fig. \ref{fig6}.

Some of the solitons, viz. Figs. \ref{fig2}(b),  \ref{fig2}(e), 
\ref{fig4}(b)-(c), and \ref{fig5}(b),  presented in this paper are excited metastable states trapped in a local minimum of energy. Their existence, stability, and eventual decay are governed by an interplay between energetic (thermodynamic) considerations and the evolution of perturbations (dynamical stability).  The imaginary-time propagation procedure has been adapted successfully for these metastable states, starting with a trial wave function with a definite symmetry property. In addition,   these solitons are dynamically stable, as verified by the real-time propagation procedure.   The stability of these metastable states in both real- and imaginary-time procedures generally may assure the thermodynamic stability of these states, but this is not guaranteed.
Being excited states, they could be thermodynamically unstable and  suffer energetic decay to the ground state, with a global minimum of energy, and phonons. A full analysis of the thermodynamic stability of these states is beyond the scope of this paper.

 We believe that the present study could motivate experiments in  the search of the novel solitons in an SO-coupled  pseudo spin-half  and spin-one dipolar spinor BEC. 
  The present solitons are dynamically robust and deserve   further  theoretical and experimental  investigations.

\section*{ACKNOWLEDGMENT}

The author  
 acknowledges support by the Conselho Nacional de Desenvolvimento  Científico e Technológico  (Brazil) grant  303885/2024-6.

 \section*{DATA AVAILABILITY}
The data that support the findings of this article are not
publicly available upon publication because it is not technically feasible and/or the cost of preparing, depositing, and
hosting the data would be prohibitive within the terms of this
research project. The data are available from the authors upon
reasonable request.

\end{document}